\documentclass[english,preprint]{aastex}
\usepackage{lmodern}
\usepackage[T1]{fontenc}
\usepackage[latin9]{inputenc}
\usepackage{color}
\usepackage{booktabs}
\usepackage{amssymb}
\usepackage{graphicx}

\makeatletter

\providecommand{\tabularnewline}{\\}
\newcommand{\lyxdot}{.}

\usepackage{amsmath}

\makeatother

\usepackage{babel}
\begin{document}

\title{Rapid Change of Field Line Connectivity and Reconnection in Stochastic
Magnetic Fields}

\author{Yi-Min Huang \altaffilmark{1,2,3,4} and A. Bhattacharjee\altaffilmark{1,2,3,4}}

\affil{Department of Astrophysical Sciences and Princeton Plasma Physics
Laboratory, Princeton University, Princeton, New Jersey 08543, USA}

\author{Allen H. Boozer\altaffilmark{1}}

\affil{Department of Applied Physics and Applied Mathematics, Columbia University,
New York, New York 10027, USA}

\altaffiltext{1}{Max Planck/Princeton Center for Plasma Physics}

\altaffiltext{2}{Princeton Center for Heliospheric Physics}

\altaffiltext{3}{Center for Magnetic Self-Organization in Laboratory and Astrophysical Plasmas}

\altaffiltext{4}{Center for Integrated Computation and Analysis of Reconnection and Turbulence}
\begin{abstract}
Magnetic fields without a direction of continuous symmetry have the 
generic feature that neighboring field lines exponentiate away from 
each other and become stochastic, hence the ideal constraint of 
preserving magnetic field line connectivity becomes exponentially 
sensitive to small deviations from ideal Ohm's law. The idea of breaking 
field line connectivity by stochasticity as a mechanism for fast 
reconnection is tested with numerical simulations based on reduced 
magnetohydrodynamics equations with a strong guide field line-tied to 
two perfectly conducting end plates. Starting from an ideally stable 
force-free equilibrium, the system is allowed to undergo resistive 
relaxation. Two distinct phases are found in the process of resistive 
relaxation. During the quasi-static phase, rapid change of field line 
connectivity and strong induced flow are found in regions of high field 
line exponentiation. However, although the field line connectivity of 
individual field lines can change rapidly, the overall pattern of field 
line mapping appears to deform gradually. From this perspective, field 
line exponentiation appears to cause enhanced diffusion rather than 
reconnection. In some cases, resistive quasi-static evolution can cause 
the ideally stable initial equilibrium to cross a stability threshold, 
leading to formation of intense current filaments and rapid change of 
field line mapping into a qualitatively different pattern. It is in this 
onset phase that the change of field line connectivity is more 
appropriately designated as magnetic reconnection. Our results show that 
rapid change of field line connectivity appears to be a necessary, 
but not a sufficient condition for fast reconnection.
\end{abstract}

\section{Introduction}

Magnetic reconnection is a fundamental process in astrophysical, space,
and laboratory plasmas, which can change the topology of magnetic
field lines, release magnetic energy, and accelerate electrons and
ions \citep{Biskamp2000,PriestF2000,ZweibelY2009,YamadaKJ2010}. It
is generally believed that magnetic reconnection is the underlying
mechanism that powers explosive events such as solar flares, coronal
mass ejections, geomagnetic substorms, and sawtooth crashes in fusion
devices. 

Much of the literature on magnetic reconnection focuses on two-dimensional
(2D) problems, i.e. when the whole reconnection process only depends
on two spatial coordinates under a proper coordinate system, for reasons
of analytic and numerical simplicity. At this time, theories of 2D
magnetic reconnection are highly developed and fairly well-understood.
Magnetic reconnection in 2D is known to take place at an X-point (or
X-line), where magnetic field line separatrices intersect; across
separatrices the field line mapping is discontinuous. As a magnetic
field line is carried across a separatrix by plasma flow, the velocity
field that preserves magnetic field line connectivity (i.e. the velocity
field carries a field line at one time to another field line at a
later time) diverges at the X-point \citep{PriestHP2003}; that is,
the field line is cut and rejoined with another field line at the
X-point.

Real world magnetic reconnection of course takes place in three dimensions
(3D). Compared with 2D problems, magnetic reconnection in 3D is much
less well-understood {[}see, e.g. \citet{Pontin2011} for a recent
review, and the references therein.{]} Even the concept of magnetic
reconnection can sometimes become ambiguous, especially in the absence
of magnetic null points or closed magnetic field lines. In these situations,
all magnetic field lines are topologically equivalent, and a continuous
velocity field that preserves magnetic field line connectivity can
always be defined \citep{Greene1993}. Where, then, does magnetic
reconnection take place? Quasi-separatrix layers (QSLs), defined as
regions where the field line mapping has large gradient \citep{PriestD1995,DemoulinHPM1996,Titov2007},
have been suggested as likely locations for reconnection to take place.
Although field line velocity can always be defined in a 3D magnetic
field without null points or closed field lines, the field line velocity
can differ significantly from the plasma velocity around QSLs. The
concept of QSLs have been widely employed in analyzing 3D magnetic
reconnection in recent numerical simulations and laboratory experiments
\citep{LawrenceG2009,RichardsonF2012,GekelmanVDV2014,FinnBDZ2014}. 

In a magnetic field with no ignorable coordinates, magnetic field
lines generically exponentiate apart from each other and become stochastic.
Such a generic condition usually occurs in space and astrophysical
plasmas as well as in a number of contexts in laboratory plasmas (including
toroidal fusion plasmas) in the presence of multiple tearing modes.
Also, global simulation codes for space weather studies often exhibit
magnetic turbulence (at fluid as well as kinetic levels) which produces
field-line exponentiation. That means that if we follow two field
lines initially separated by an infinitesimal distance $\delta r(0)$,
in most cases the separation grows exponentially as $\delta r(\ell)=e^{\sigma(\ell)}\delta r(0)$,
where $\ell$ is the distance along the field line, and $\sigma(\ell)$
is an overall (but in general not monotonically) increasing function
over distance. This fact is analogous to the well-known butterfly
effect in the theory of deterministic dynamical systems as a sensitive
dependence on initial conditions. \citet{Boozer2012a,Boozer2012,Boozer2013}
has recently argued that under the condition of large field line exponentiation,
an exponentially small non-ideal effect will completely scramble the
field line mapping, and potentially lead to fast reconnection. Although
the notion of large field line exponentiation shares common ground
with the concept of QSLs, an important difference is that the condition
for neighboring field line exponentiation is valid essentially for
all field lines, therefore regions of large field line exponentiation
do not necessarily concentrate in layers and can even be volume filling.

The motivation of this study is to test this scenario proposed by
Boozer with numerical experiments as clean as possible. To this end
we consider relaxation of force-free equilibria in a region bounded
by two conducting plates under the effect of a finite but small resistivity.
Without resistivity, the field line mapping from one end to another
is preserved. With a finite resistivity, the field line mapping evolves
as time proceeds. We find that in general the field line connectivity
changes rapidly in regions with large neighboring field line exponentiation.
However, not all rapid changes of field line connectivity are associated
with what is usually regarded as reconnection. We find two distinct
phases of evolution --- one is the quasi-static phase and the other
the onset phase. The onset phase is triggered by some magnetohydrodynamic
(MHD) instability. We find that change of field line connectivity
during the quasi-static phase is more properly attributed to enhanced
magnetic diffusion, whereas typical hallmarks of reconnection are
found during the onset phase.

This paper is organized as follows. In Section \ref{sec:Model-and-Simulation}
we introduce the reduced magnetohydrodynamics (RMHD) model and simulation
setup. In Section \ref{sec:Diagnostics}, we introduce the details
of diagnostics used to analyze the simulations. Section \ref{sec:Simulation-Results}
presents two sets of simulations. The first set is governed by resistive
quasi-static evolution, and the second set features the onset phase.
Section \ref{sec:Discussion} discusses important issues that arise
from the simulation results, including the governing equations for
resistive quasi-static evolution, the cause of the onset phase, and
some measures to quantify the distinction between the two phases.
Finally, we summarize the key findings and conclude in Section \ref{sec:Summary-and-Conclusion}.
{In Appendix \ref{sec:Construction-of-Force-Free},
we explain the procedure of constructing the initial RMHD force-free
equilibrium in detail, and make comparison with the corresponding
problem in full MHD.}

\section{Model and Simulation Setup \label{sec:Model-and-Simulation}}

We employ the standard reduced magnetohydrodynamics (RMHD) model \citep{KadomtsevP1974,Strauss1976,VanBallegooijen1985},
which is valid in the presence of a strong uniform guide field, under
the assumption that spatial length scales are much longer along the
direction of the guide field than in directions perpendicular to the
guide field. The governing equations can be written as:

\begin{equation}
\partial_{t}\Omega+[\phi,\Omega]=\partial_{z}J+[A,J]+\nu\nabla_{\perp}^{2}\Omega-\lambda\Omega,\label{eq:RMHD-momentum}
\end{equation}
\begin{equation}
\partial_{t}A+[\phi,A]=\partial_{z}\phi+\eta\nabla_{\perp}^{2}A.\label{eq:RMHD-faraday}
\end{equation}
Here we assume the guide field to be along the $z$ direction, with
the strength normalized to unity. The magnetic field $\mathbf{B}$
and the plasma velocity $\mathbf{u}$ are expressed in terms of the
flux function $A$ and the stream function $\phi$ through the relations
$\mathbf{B}=\mathbf{\hat{z}}+\nabla_{\perp}A\times\mathbf{\hat{z}}$
and $\mathbf{u}=\nabla_{\perp}\phi\times\mathbf{\hat{z}}$; and $\Omega\equiv-\nabla_{\perp}^{2}\phi$
and $J\equiv-\nabla_{\perp}^{2}A$ are vorticity and electric current
along the $z$ direction, respectively. The Poisson bracket is defined
as $\left[f,g\right]=\partial_{y}f\partial_{x}g-\partial_{x}f\partial_{y}g$.
Dissipations are introduced by including resistivity $\eta$, viscosity
$\nu$, and a friction coefficient $\lambda$. We assume that the
system is bounded in the $z$ direction by two conducting end plates
at $z=0$ and $z=L$, and is a $1\times1$ box with doubly periodic
boundary condition in the $x-y$ plane. RMHD models are often used
in analytic and numerical studies of Parker's coronal heating model
\citep{VanBallegooijen1985,StraussO1988,LongcopeS1994a,LongcopeS1994b,NgB1998,DmitrukG1999,DmitrukGM2003,RappazzoVED2007,RappazzoVED2008,NgB2008,NgLB2012,RappazzoP2013}.
However, unlike simulations of Parker's model, here we impose no boundary
flow, i.e. the boundary conditions are $\phi=0$ at $z=0$ and $z=L$.
The governing equations are solved with a reduced version of the compressible
MHD code DEBS \citep{SchnackBMHCN1986}. The $x-y$ plane is discretized
with a pseudospectral method, and the $z$ direction is discretized
with a finite difference method, where $\phi$ and $A$ reside on
staggered grids. The timestepping is carried out with a semi-implicit,
predictor-corrector leapfrog method, where $\phi$ and $A$ are staggered
in time as well.

In this model, resistivity provides the mechanism that breaks the
ideal constraint of frozen field lines, whereas viscosity and friction
are introduced to damp Alfv\'en waves that will be generated when
the system is not in force equilibrium, e.g. due to reconnection.
Undamped Alfv\'en waves bouncing back and forth between the two conducting
end plates may lead to fine structures through distortion of wave
packets and phase mixing, especially in regions where the magnetic
field is highly sheared \citep{SimilonS1989}. Although this phenomenon
can be very important in wave dissipation and heating, it further
complicates the problem and may lead to numerical difficulties. For
these reasons, we opt for damping out Alfv\'en waves in order to
to make the simulations as clean as possible.

A force-free equilibrium in RMHD satisfies the condition $\partial_{z}J+\left[A,J\right]=\mathbf{B}\cdot\nabla J=0.$
Formally, this equation is isomorphic to the 2D Euler's equation of
incompressible fluid, if we identify $z$ with a time variable, and
$A$ with the stream function. Therefore, if $A$ is specified on
a $x-y$ plane, we can construct the corresponding force-free equilibrium
by integrating $\mathbf{B}\cdot\nabla J=0$ along $z$ {(see
Appendix \ref{sec:Construction-of-Force-Free} for details of numerical
implementation)}. In our simulations, we construct the initial conditions
by specifying the flux function at the mid-plane $A|_{z=L/2}$ as
a superposition of Fourier harmonics 
\begin{equation}
A|_{z=L/2}=\sum_{m,n}a_{mn}\exp(i\pi(mx+ny)),\label{eq:A0}
\end{equation}
where the coefficients $a_{mn}$ are complex random numbers whose
magnitudes $\left|a_{mn}\right|$ on average scale as $\left|a_{mn}\right|\sim\left(m^{2}+n^{2}\right)^{-2}$.
Only mode numbers $m$, $n$ within the range $2\le\sqrt{m^{2}+n^{2}}\le5$
are used. In addition, the coefficients $a_{mn}$ must satisfy the
reality condition $a_{mn}=a_{-m-n}^{*}$. Because Euler's equation
conserves energy, the equilibrium satisfies 
\begin{equation}
\frac{d}{dz}\int\frac{1}{2}B_{\perp}^{2}dxdy=0,\label{eq:energy_conservation}
\end{equation}
where $\mathbf{B}_{\perp}=\nabla_{\perp}A\times\mathbf{\hat{z}}$
is the magnetic field component on the $x-y$ plane. 

In this study, first a set of Fourier coefficients $a_{mn}$ are randomly
chosen. Subsequently, they are rescaled to a given $E_{\perp}\equiv\int\frac{1}{2}B_{\perp}^{2}dxdy$,
which is used as a free parameter to prescribe the initial equilibrium.
We set the system length to a fixed value $L=10$. This value of system
length is arbitrary and does not affect the results, because the RMHD
model is scale-invariant. That is, if $\phi(\mathbf{x_{\perp}},z,t)$
and $A(\mathbf{x_{\perp}},z,t)$ form a solution of RMHD with dissipation
coefficients $\nu$, $\eta$, and $\lambda$, then for an arbitrary
rescale factor $\zeta$, the rescaled solutions $\phi\to\zeta\phi(\mathbf{x_{\perp}},\zeta z,\zeta t)$
and $A\to\zeta A(\mathbf{x_{\perp}},\zeta z,\zeta t)$ satisfy RMHD
with dissipation coefficients $\nu\to\zeta\nu$, $\eta\to\zeta\eta$,
and $\lambda\to\zeta\lambda$. Therefore, a solution with respect
to a given system length $L$ can always be rescaled to another solution
with a system length $L'=\zeta L$.

\begin{figure}
\begin{centering}
\includegraphics[width=6in]{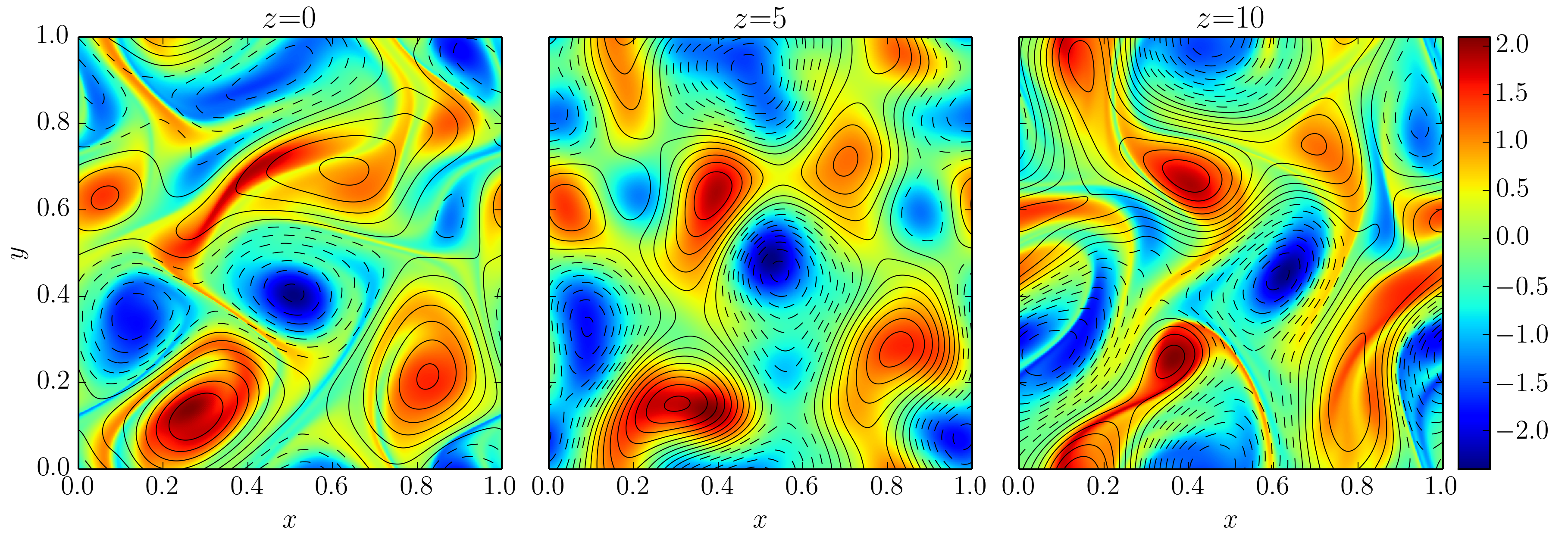}
\par\end{centering}

\protect\caption{Slices of initial current density $J$ profile, overlaid with contours
of the flux function $A$, on $x-y$ planes at $z=0$, $z=5$, and
$z=10$, for the case $E_{\perp}=1.4\times10^{-3}$. \label{fig:current_profile}}

\end{figure}

Figure \ref{fig:current_profile} show slices of initial current density
profile, overlaid with contours of the flux function $A$, on $x-y$
planes at $z=0$, $z=5$, and $z=10$, for the case $E_{\perp}=1.4\times10^{-3}$.
Because Equation (\ref{eq:A0}) only contains long wavelength Fourier
harmonics, the current density $J$ is smooth at $z=5$. However,
as we integrate $\mathbf{B}\cdot\nabla J=0$ along both $\pm z$ directions,
constant-$J$ contours become increasingly stretched and develop fine
structures due to field line exponentiation. If the neighboring field
lines exponentiate by a factor of $e^{\sigma}$, $\left|\nabla J\right|$
can be amplified by a factor up to $e^{\sigma}$. In the present case,
although $J$ is of the order unity in the whole domain, $\left|\nabla J\right|$
clearly becomes large at $z=0$ and $z=10$.

Once the initial equilibrium is constructed, we study resistive relaxation
under different values of resistivity $\eta$. We fix the viscosity
$\nu=10^{-6}$ and the friction coefficient $\lambda=0.1$. Through
numerical experiments, we find that these values are sufficient to
effectively damp out Alfv\'en wave in a few Alfv\'en transit times.

\section{Diagnostics \label{sec:Diagnostics}}

Neighboring field line exponentiation is characterized by the exponent
$\sigma$, which can be calculated by simultaneously integrating the
equation for field line

\begin{equation}
\frac{d\mathbf{x}_{\perp}}{dz}=\mathbf{B_{\perp}}\label{eq:field_line}
\end{equation}
 and the equation for infinitesimal separation $\delta\mathbf{x}_{\perp}$
between neighboring field lines: 
\begin{equation}
\frac{d\delta\mathbf{x}_{\perp}}{dz}=\delta\mathbf{x}_{\perp}\cdot\nabla_{\perp}\mathbf{B}_{\perp}.\label{eq:separation}
\end{equation}
In terms of the flux function $A$, Equation (\ref{eq:field_line})
can be written as 
\begin{equation}
\frac{d}{dz}\left[\begin{array}{c}
x\\
y
\end{array}\right]=\left[\begin{array}{c}
\partial_{y}A\\
-\partial_{x}A
\end{array}\right].\label{eq:field_line_hamiltonian}
\end{equation}
Hence, the flux function $A$ is the Hamiltonian for field lines,
if we identify $z$ with a time variable. Equation (\ref{eq:separation})
can be written in matrix form as: 
\begin{equation}
\frac{d}{dz}\left[\begin{array}{c}
\delta x\\
\delta y
\end{array}\right]=\left[\begin{array}{cc}
\partial_{x}B_{x} & \partial_{y}B_{x}\\
\partial_{x}B_{y} & \partial_{y}B_{y}
\end{array}\right]\left[\begin{array}{c}
\delta x\\
\delta y
\end{array}\right]\equiv M(z)\left[\begin{array}{c}
\delta x\\
\delta y
\end{array}\right],\label{eq:matrix_form}
\end{equation}
which can be formally solved as 
\begin{equation}
\left[\begin{array}{c}
\delta x\\
\delta y
\end{array}\right]=\exp\left(\int_{0}^{z}M(z')dz'\right)\left[\begin{array}{c}
\delta x\\
\delta y
\end{array}\right]_{z=0}\equiv N(z)\left[\begin{array}{c}
\delta x\\
\delta y
\end{array}\right]_{z=0},\label{eq:solution}
\end{equation}
where the integration is carried out along the field line. However,
direct numerical calculation of $\exp\left(\int_{0}^{z}M(z')dz'\right)$
is prone to numerical instability. Therefore, in practice the matrix
$N(z)$ is calculated by integrating the equation 
\begin{equation}
\frac{dN}{dz}=MN,\label{eq:matrix_eq}
\end{equation}
with the initial condition 
\begin{equation}
N|_{z=0}=\left[\begin{array}{cc}
1 & 0\\
0 & 1
\end{array}\right].\label{eq:initial_cond}
\end{equation}
Because the field line mapping in RMHD preserves area, the singular
value decomposition (SVD) \citep{TrefethenB1997} of $N(z)$ is of
the form 
\begin{equation}
N(z)=U(z)\left[\begin{array}{cc}
e^{\sigma(z)} & 0\\
0 & e^{-\sigma(z)}
\end{array}\right]V^{T}(z),\label{eq:SVD}
\end{equation}
where both $U(z)$ and $V(z)$ are unitary matrices. The factor $e^{\sigma(z)}$
gives the maximum exponentiation we could have for two neighboring
field lines. A geometrical interpretation of the exponent $\sigma(z)$
is that, if we follow a thin flux tube starting with a circular cross
section of radius $\delta r$ at $z=0$, the cross section becomes
an ellipse at $z>0$, with $\delta r_{max}=e^{\sigma(z)}\delta r$
and $\delta r_{min}=e^{-\sigma(z)}\delta r$ as the major and minor
radii, respectively. This interpretation bears resemblance to the
squashing factor $Q$ often employed in the definition of QSLs \citep{Titov2007},
which is defined as 
\begin{equation}
Q=\frac{\delta r_{max}}{\delta r_{min}}+\frac{\delta r_{min}}{\delta r_{max}}\label{eq:squashing}
\end{equation}
measured at the top plate. In the limit $\sigma(L)\gg1$, the exponent
$\sigma(L)$ is related to the squashing factor $Q$ by the relation
$Q\simeq e^{2\sigma(L)}$. However, note that neighboring field line
exponentiation is a more general concept, in the sense that the squashing
factor $Q$ depends only on the field line mapping from one end plate
to another, but not on the field line separation that occurs in the
region between the two plates.

In this study, individual field lines are treated as fundamental entities,
labeled by their footpoints at the bottom plate. Along each field
line, we calculate the following quantities: 
\begin{enumerate}
\item The maximum value of the exponent $\sigma_{max}\equiv\underset{0\le z\le L}{\max}\sigma(z)$.
This provides information of the locations where the neighboring field
lines strongly exponentiate apart.
\item Drift of footpoints of magnetic field lines at the top plate between
two consecutive snapshots separated by one Alfv\'en transit time
($=10$ in the simulation) when the footpoints at the bottom plate
is held fixed. Without non-ideal effects, the footpoint drift should
be identically zero.
\item Kinetic energy density $\int_{0}^{L}\frac{1}{2}\left|\nabla_{\perp}\phi\right|^{2}dz$.
This provides information of where the energy conversion from magnetic
field to kinetic energy occurs.
\item Parallel electric voltage $\int\mathbf{E}\cdot d\mathbf{l}=\int_{0}^{L}\eta Jdz$.
This is often employed as a proxy for reconnection rate in 3D \citep{SchindlerHB1988,HesseS1988,HesseFB2005}. 
\end{enumerate}
By using field lines as the fundamental entities, we effectively reduce
the 3D information to 2D. We typically trace $400\times400$ field
lines arranged in a uniform array at the bottom plate in each snapshot.

\begin{figure}
\begin{centering}
\includegraphics[scale=0.5]{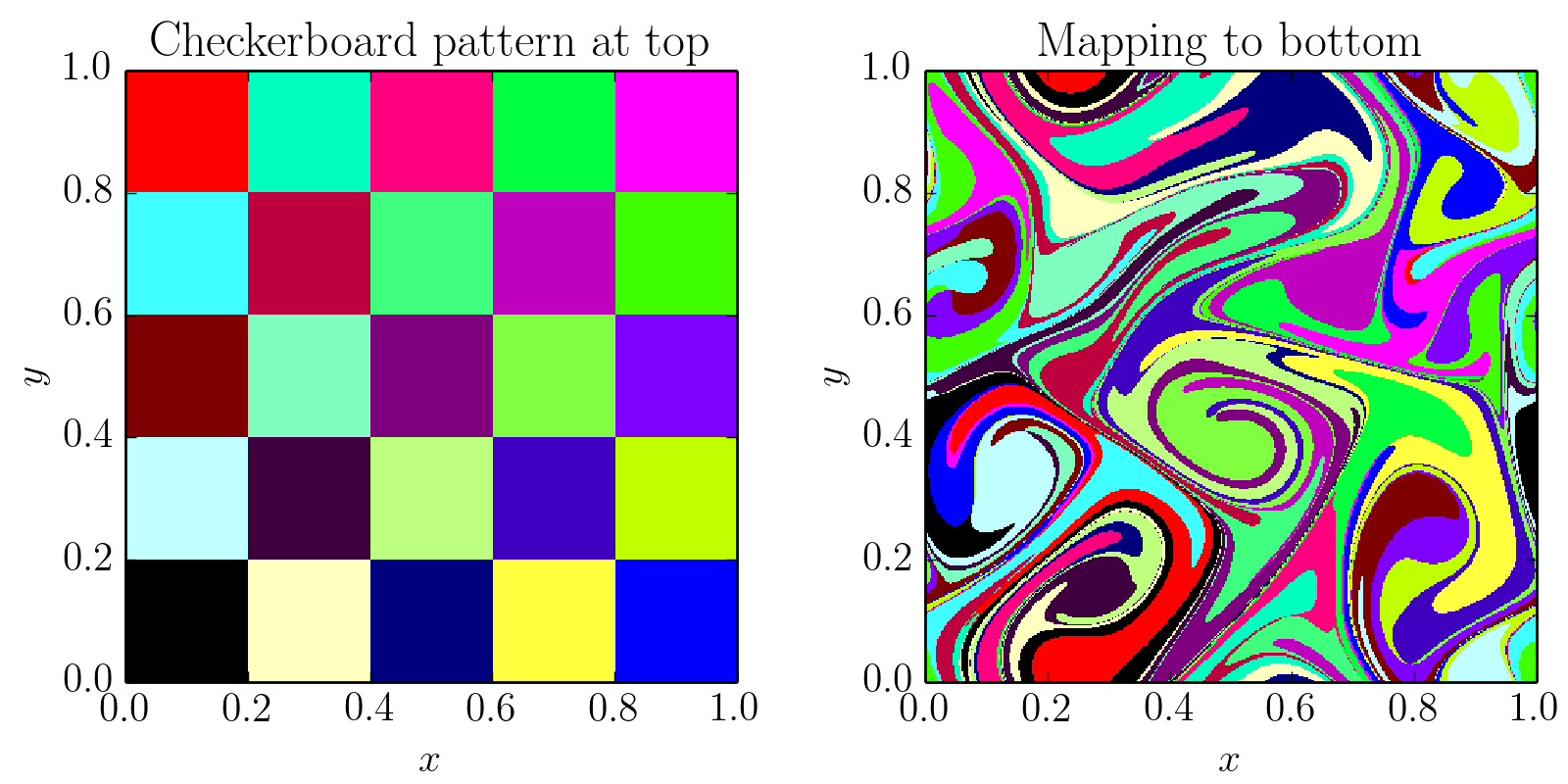}
\par\end{centering}

\protect\caption{Visualization of footpoint mapping between two end plates, for the
initial condition with $E_{\perp}=1.4\times10^{-3}$. Left panel shows
a $5\times5$ checkerboard pattern at the top plate; the right panel
shows the pattern at the bottom plate, when the checkerboard pattern
is pulled back by the footpoint mapping. \label{fig:Visualization-of-mapping}}

\end{figure}

We visualize the field line mapping between two plates as follows.
First a checkerboard pattern is laid out at the top plate, which is
then pulled back to the bottom plate by the field line mapping. The
resulting pattern at the bottom plate gives a visualization of the
field line mapping. Figure \ref{fig:Visualization-of-mapping} shows
an example of such visualization for the initial condition with $E_{\perp}=1.4\times10^{-3}$.
Here we have made use of the doubly periodic boundary condition in
defining field line mapping between the two plates, such that the
top plate of the simulation box is exactly mapped to the bottom plate.
As a result of neighboring field line exponentiation, the checkerboard
pattern becomes highly distorted when pulled back to the bottom plate.

\section{Simulation Results\label{sec:Simulation-Results}}

Two sets of simulations have been carried out for this study, with
$E_{\perp}=1.25\times10^{-3}$ and $E_{\perp}=1.4\times10^{-3}$,
respectively. Both initial conditions are ideally stable; i.e. they
remain unchanged for a long time when $\eta$ is set to zero. The
two initial conditions are subjected to resistive relaxation with
different $\eta$. Parameters of the reported simulations in this
paper are listed in Table \ref{tab:Parameters}. Convergence test
has been carried out for selected runs. {Specifically,
Run B2 and Run B3 have been tested with a lower resolution $512^{3}$,
and the results are essentially the same as the higher resolution
runs presented in the paper.}

\begin{table}[t]
\begin{centering}
\begin{tabular}{cccc}
\toprule 
Run & $E_{\perp}$ & $\eta$ & Resolution\tabularnewline
\midrule 
A1 & $1.25\times10^{-3}$ & $10^{-5}$ & $512^{3}$\tabularnewline
A2 & $1.25\times10^{-3}$ & $5\times10^{-6}$ & $512^{3}$\tabularnewline
A3 & $1.25\times10^{-3}$ & $2.5\times10^{-6}$ & $512^{3}$\tabularnewline
A4 & $1.25\times10^{-3}$ & $10^{-6}$ & $512^{3}$\tabularnewline
B1 & $1.4\times10^{-3}$ & $10^{-5}$ & $512^{3}$\tabularnewline
B2 & $1.4\times10^{-3}$ & $5\times10^{-6}$ & $768^{3}$\tabularnewline
B3 & $1.4\times10^{-3}$ & $2.5\times10^{-6}$ & $768^{3}$\tabularnewline
\bottomrule
\end{tabular}
\par\end{centering}

\protect\caption{Simulation parameters of the Runs reported in this paper. The viscosity
$\nu$ and friction coefficient $\lambda$ are kept constant with
$\nu=10^{-6}$ and $\lambda=0.1$. \label{tab:Parameters}}

\end{table}

\subsection{Case A: $E_{\perp}=1.25\times10^{-3}$}

\begin{figure}
\begin{centering}
\includegraphics[scale=0.5]{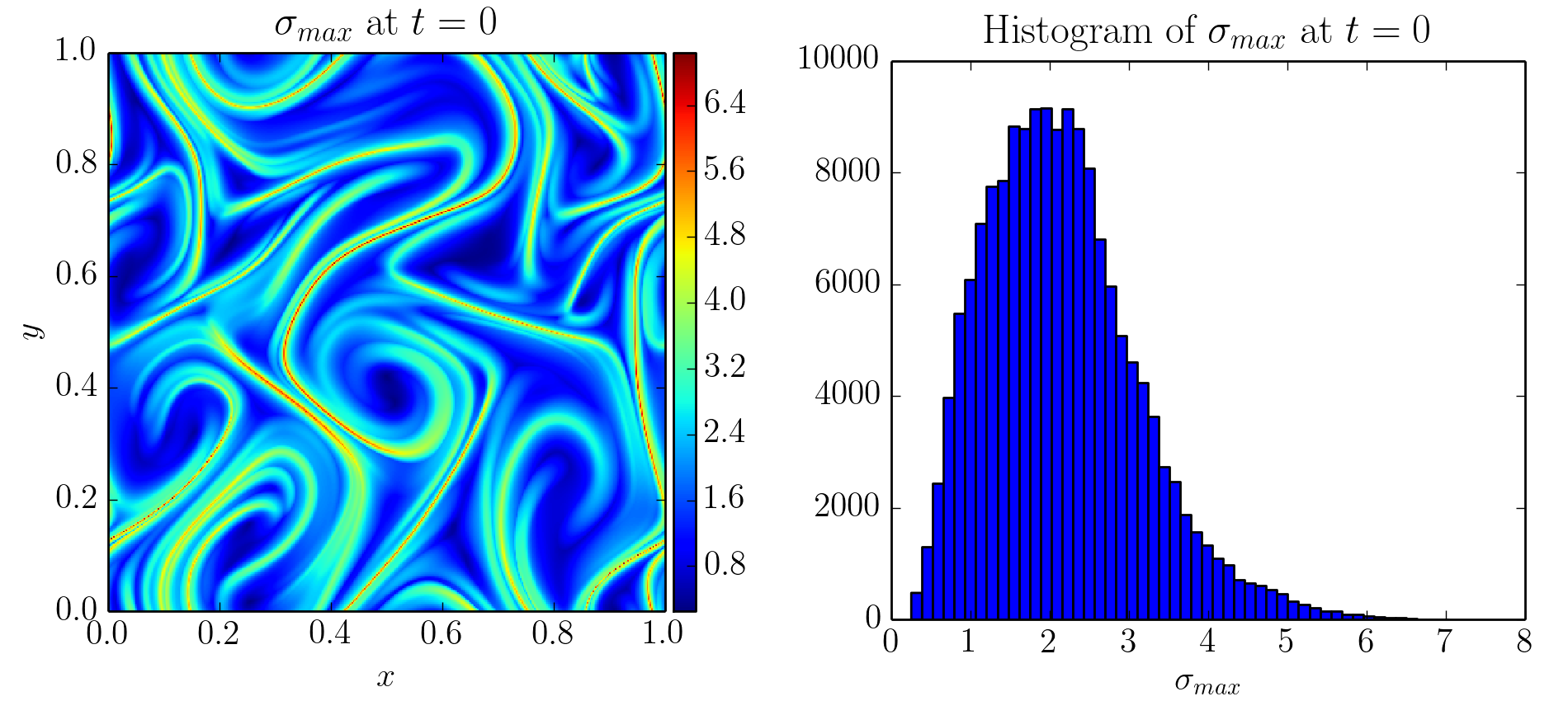}
\par\end{centering}

\protect\caption{Left panel shows the initial $\sigma_{max}$ profile for Case A with
$E_{\perp}=1.25\times10^{-3}$, labeled by the footpoints of field
lines at the bottom plate. Right panel shows the histogram of $\sigma_{max}$.\label{fig:sigma_A}}
 
\end{figure}
\begin{figure}
\begin{centering}
\includegraphics[scale=0.43]{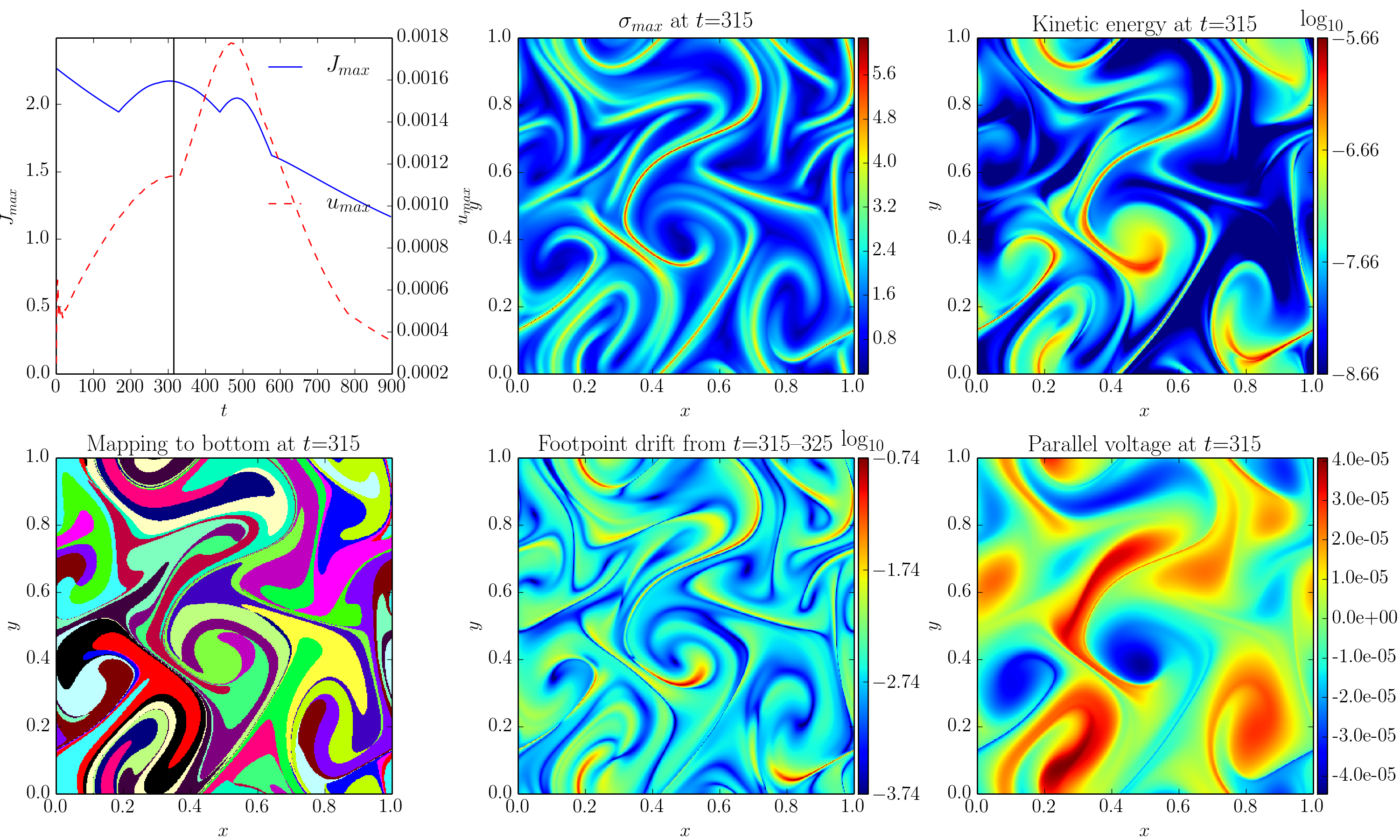}
\par\end{centering}

\protect\caption{A snapshot at $t=315$ from Run A3. Top left panel shows the time
histories of maximum current $J_{max}$ and flow velocity $u_{max}$
of the whole domain, where the vertical line indicates the time the
snapshot is taken. The remaining five panels show key diagnostics:
field line mapping, and profiles of $\sigma_{max}$, footpoint drift,
kinetic energy density, and parallel voltage (movie available online).
Note that the kinetic energy density and footpoint drift are colored
in logarithmic scales. \label{fig:CaseA-snapshot}}
\end{figure}
\begin{figure}
\begin{centering}
\includegraphics[scale=0.8]{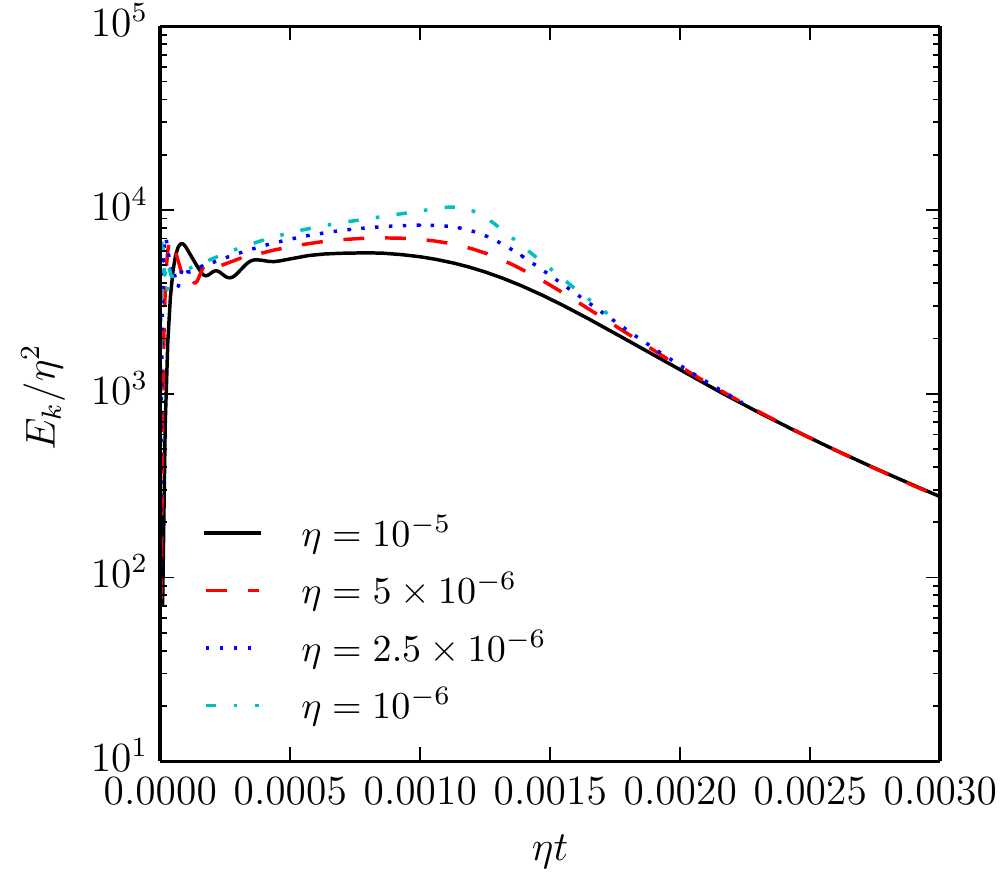}
\par\end{centering}

\protect\caption{The rescaled total kinetic energy $E_{k}/\eta^{2}$ against the rescaled
time $\eta t$ for four different $\eta$ in Case A. The four curves
approximately collapse to the same curve, except that there are slight
deviations in the middle, when $0.0005\lesssim\eta t\lesssim0.0015$.\label{fig:rescale-A} }
\end{figure}

The first set of simulations A1 -- A4 start with the initial condition
set by $E_{\perp}=1.25\times10^{-3}$. The initial profile and histogram
of $\sigma_{max}$ are shown in Figure \ref{fig:sigma_A}. As shown
in the left panel, regions with high $\sigma_{max}$ typically form
layers. The histogram in the right panel shows that for the majority
of field lines, $\sigma_{max}$ is around 2. The maximum $\sigma_{max}$
in the whole domain is approximately 7.1.

When the simulation is run with $\eta\neq0$, the magnetic field decays
resistively. Figure \ref{fig:CaseA-snapshot} shows a snapshot of
diagnostics at $t=315$ from Run A3. The top left panel shows the
time history of maximum current $J_{max}$ and flow velocity $u_{max}$
within the whole domain, where the vertical line indicates the time
the snapshot is taken. The remaining five panels show key diagnostics:
field line mapping, and profiles of $\sigma_{max}$, footpoint drift,
kinetic energy density, and parallel voltage. All the snapshots from
this Run are available online as a movie. From the time histories,
we can see that overall $J_{max}$ decays over time, whereas $u_{max}$
gradually increases in the beginning and reaches a peak at $t\simeq500$,
then gradually decays. We also observe that the $\sigma_{max}$ profile
is highly correlated with the kinetic energy density and footpoint
drift. Usually the kinetic energy density and footpoint drift tend
to be high at regions where $\sigma_{max}$ is high. The parallel
voltage remains low ($<5\times10^{-5}$) throughout the simulation
and no thin current sheets have formed.

The footpoints of individual field lines in high-$\sigma_{max}$ regions
can drift significantly (up to $\simeq0.3$) within one Alfv\'en
transit time ($=10$ in the simulation unit), assuming that the footpoints
at the bottom plate are held fixed. That gives a footpoint drift speed
up to $\simeq0.03$. In comparison, the plasma flow speed only reaches
a maximum $\simeq0.0018$, which is significantly lower than the footpoint
speed. The large deviation between plasma flow speed and field line
speed suggests possible occurrence of magnetic reconnection. However,
at the lower left panel of the movie associated with Figure \ref{fig:CaseA-snapshot},
we do not observe any sudden change in the overall pattern of footpoint
mapping, as one would expect from magnetic reconnection. Rather, the
initial convoluted field line mapping gradually untwists and becomes
simpler, and the dynamics is quiescent and uneventful. Therefore,
the rapid footpoint speed here appears to be an artifact caused by
magnetic diffusion in the presence of large gradients in the field
line mapping, rather than a manifestation of qualitative change in
the mapping itself.

Resistive relaxation of this initial condition has been carried out
for four different $\eta$. If the dynamics is solely governed by
resistive diffusion, we expect that the evolution time scale $\sim1/\eta$
and flow speed $\sim\eta$. Therefore, if we plot the rescaled total
kinetic energy $E_{k}/\eta^{2}$ against the rescaled time $\eta t$,
the curves from different runs should collapse to the same curve.
Figure \ref{fig:rescale-A} shows the rescaled curves for four different
$\eta$. And indeed the four curves approximately collapse to the
same curve, except that there are slight deviations in the middle,
when $0.0005\lesssim\eta t\lesssim0.0015$. Therefore, we conclude
that in this case the dynamics is largely governed by resistive diffusion,
rather than reconnection, even if the footpoints of individual field
lines can drift rapidly. Because the overall time scale $\sim1/\eta$,
the apparent ``drift velocity'' of footpoint is expected to scale
as $\sim\eta$.

\subsection{Case B: $E_{\perp}=1.4\times10^{-3}$}

\begin{figure}
\begin{centering}
\includegraphics[scale=0.5]{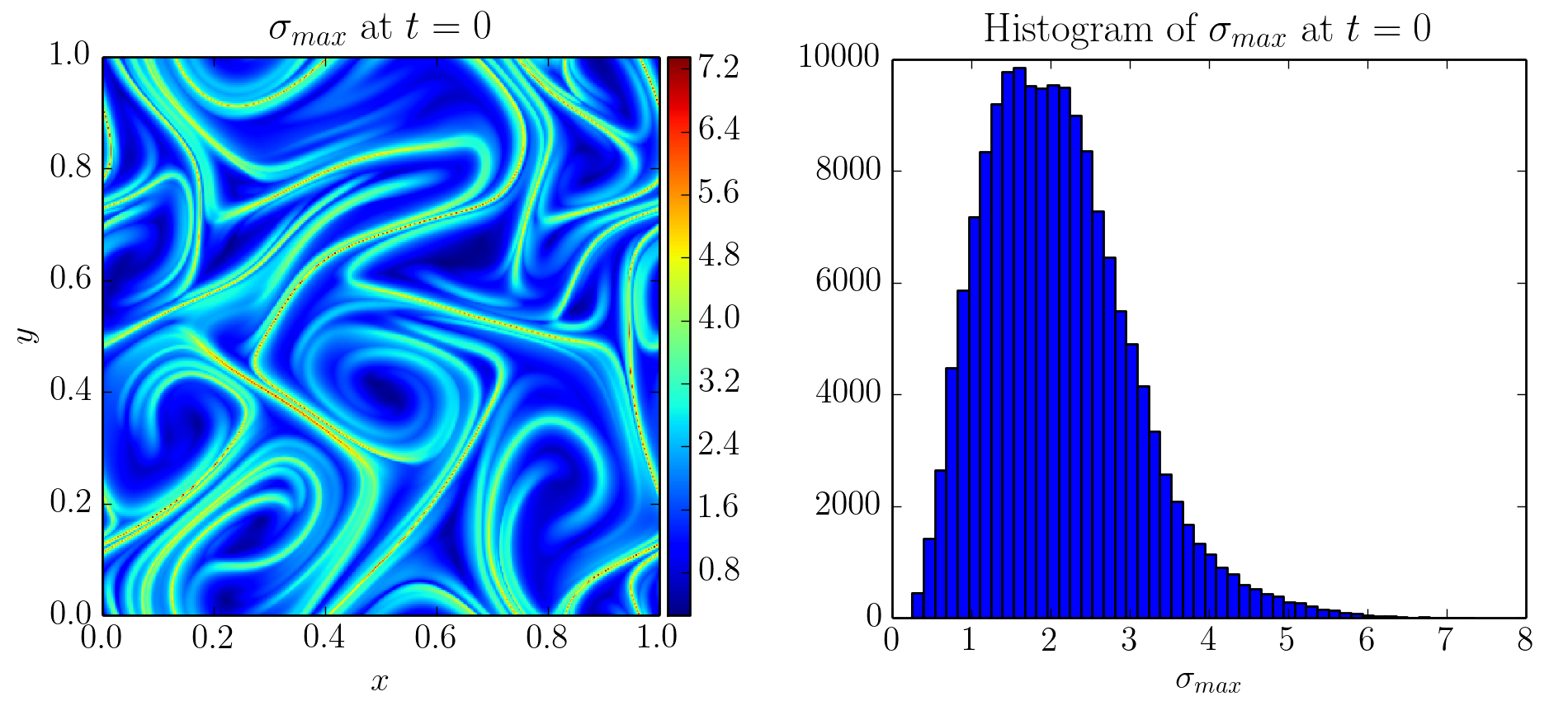}
\par\end{centering}

\protect\caption{Left panel shows the initial $\sigma_{max}$ profile for Case B with
$E_{\perp}=1.4\times10^{-3}$, labeled by the footpoints of field
lines at the bottom plate. Right panel shows the histogram of $\sigma_{max}$.\label{fig:sigma_B}}
\end{figure}
\begin{figure}
\begin{centering}
\includegraphics[scale=0.43]{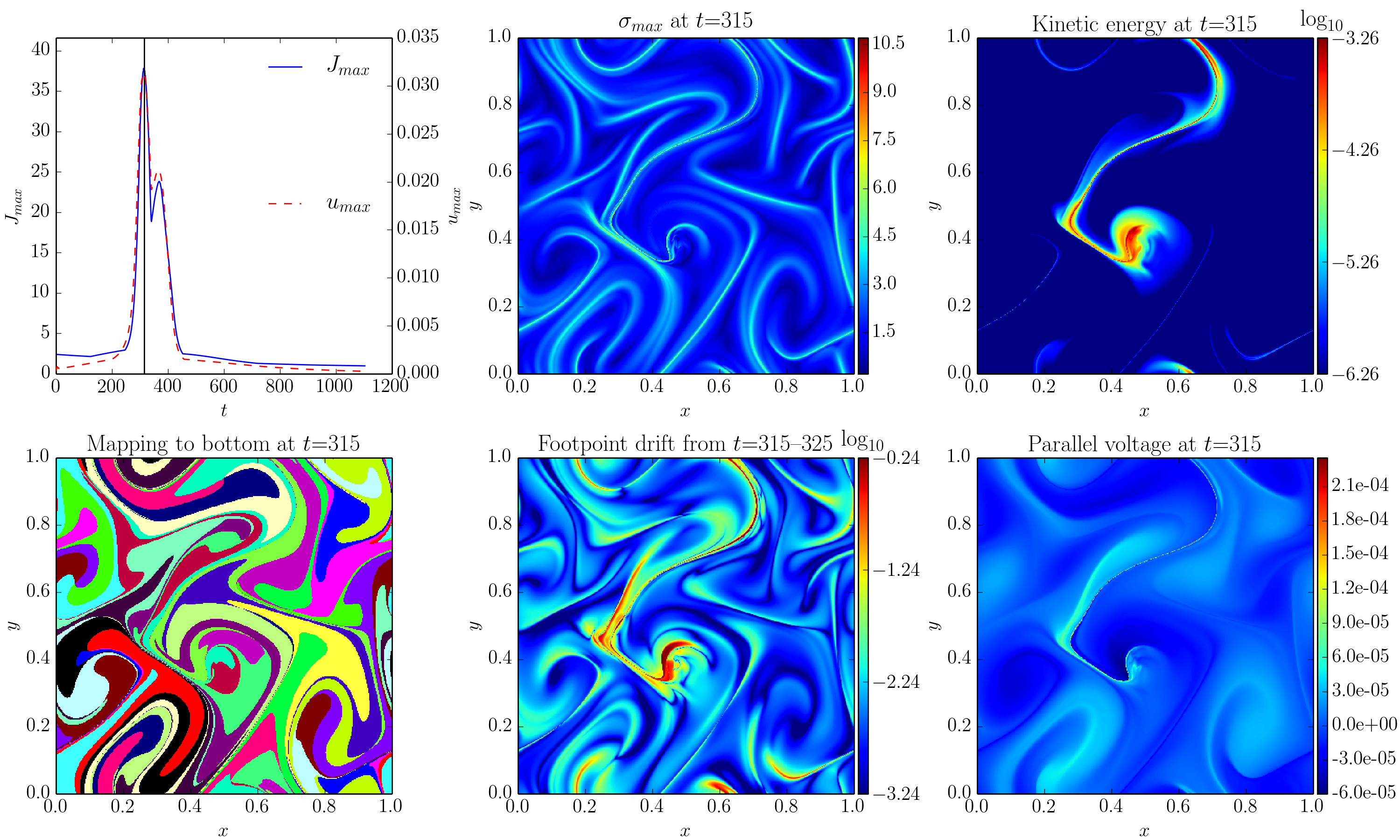}
\par\end{centering}

\protect\caption{A snapshot at $t=315$ from Run B3. Top left panel shows the time
history of maximum current $J_{max}$ and flow velocity $u_{max}$
of the whole domain, where the vertical line indicates the time the
snapshot is taken. The remaining five panels show key diagnostics:
field line mapping, and profiles of $\sigma_{max}$, footpoint drift,
kinetic energy density, and parallel voltage (movie available online).
Note that the kinetic energy density and footpoint drift are colored
in logarithmic scales.\label{fig:CaseB-snapshot} }
\end{figure}
\begin{figure}
\begin{centering}
\includegraphics[scale=0.18]{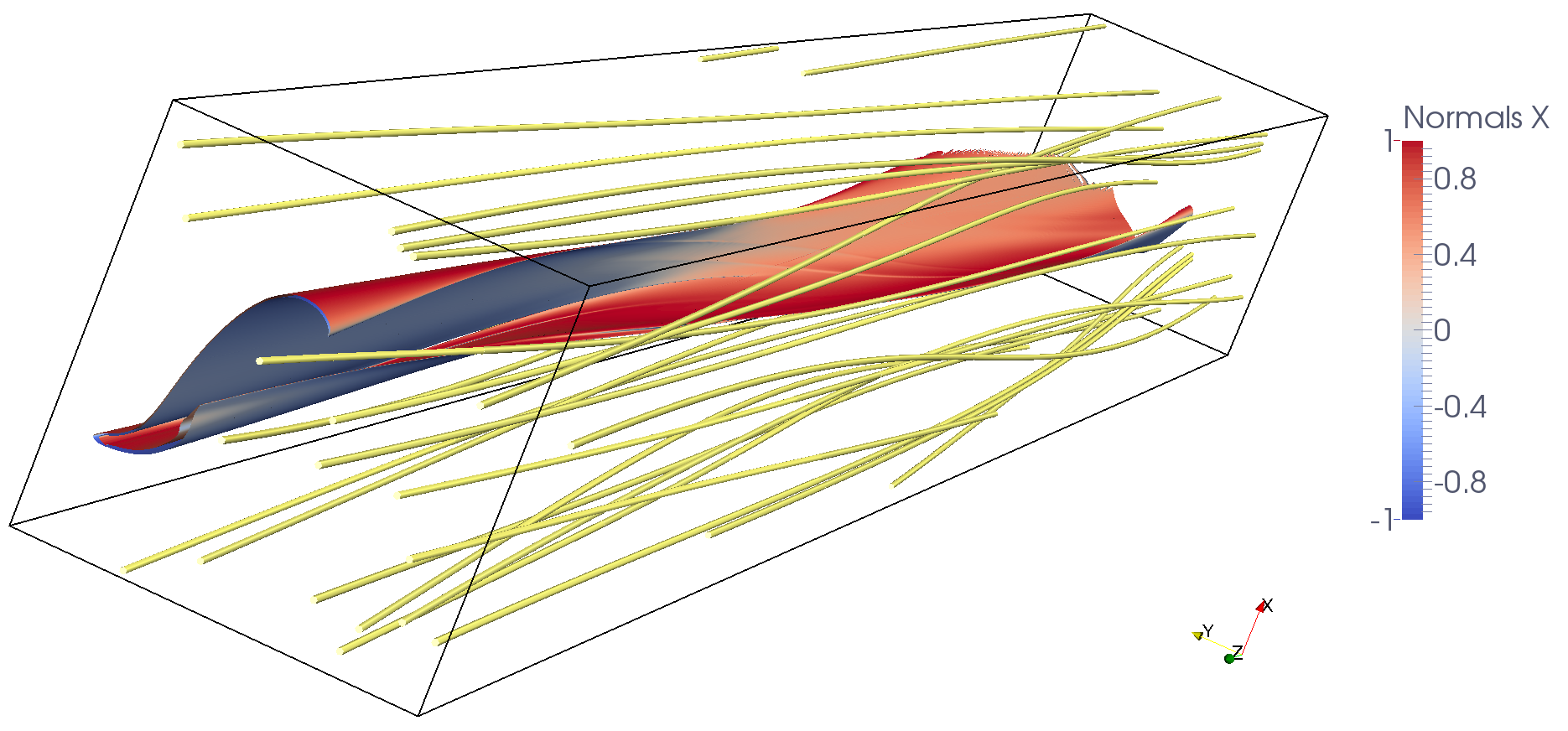}
\par\end{centering}

\protect\caption{Isosurface of the current density with $J=2$, at $t=315$ when the
current intensity peaks. The isosurface is colored according to the
$x$-component of the unit normal vector for better visualization.
Yellow lines are samples of magnetic field lines. \label{fig:Isosurface}}
\end{figure}
\begin{figure}
\begin{centering}
\includegraphics[scale=0.6]{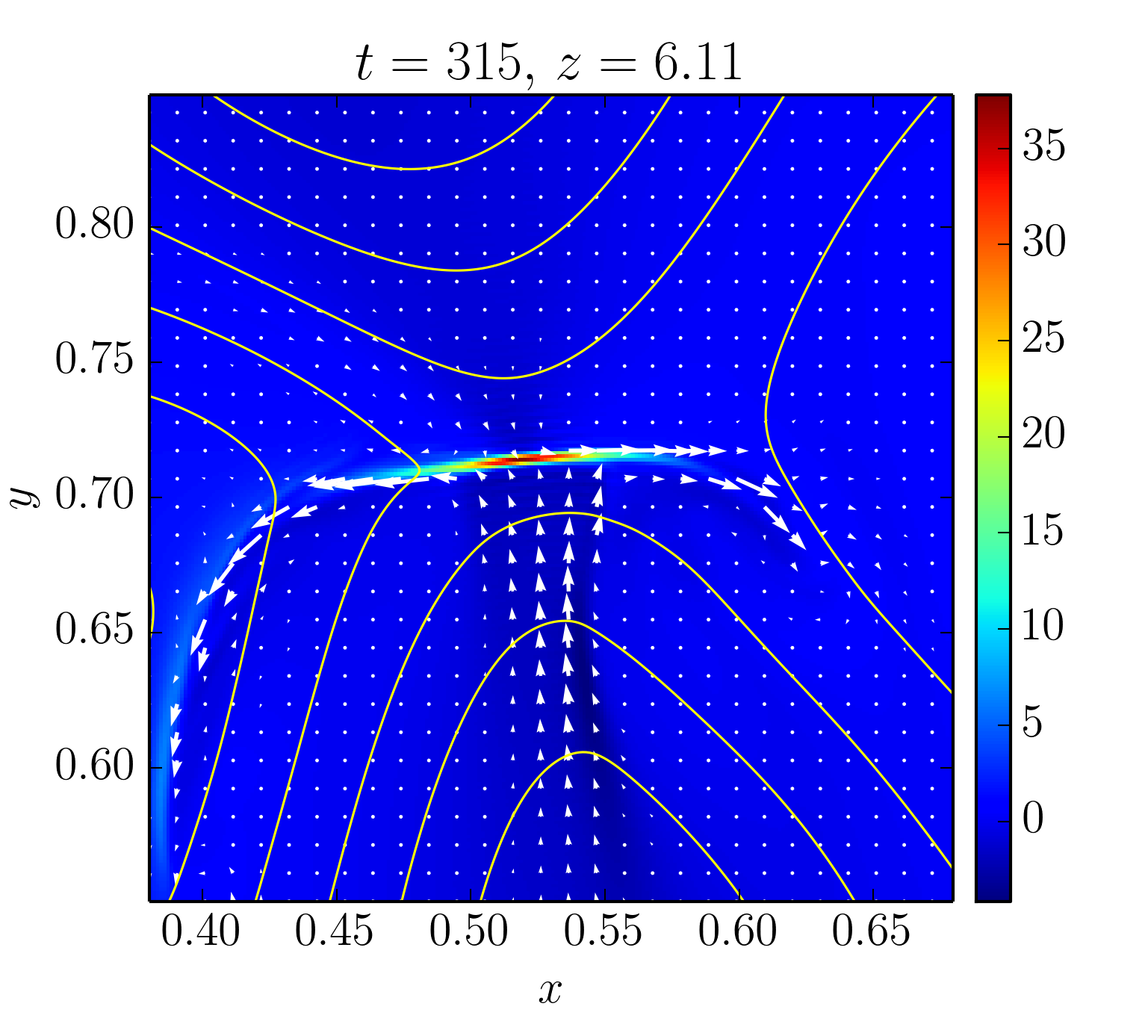}
\par\end{centering}

\protect\caption{2D slice of current density (color gradient), plasma flow (white arrows),
and contours of flux function $A$ (yellow solid lines) along $z=6.11$,
where the current density peaks at $t=315$. \label{fig:reconnection}}
\end{figure}
\begin{figure}
\begin{centering}
\includegraphics[scale=0.8]{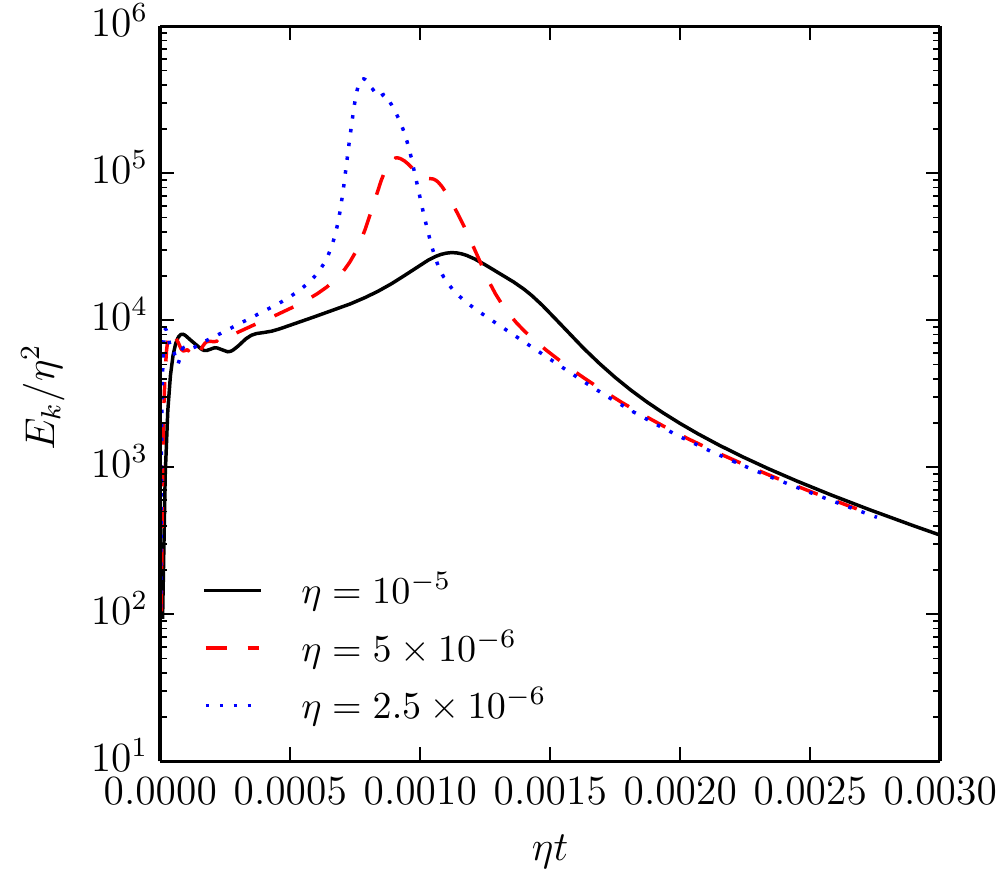}
\par\end{centering}

\protect\caption{The rescaled total kinetic energy $E_{k}/\eta^{2}$ against the rescaled
time $\eta t$ for three different $\eta$ in Case B. The thee curves
approximately collapse to the same curve during the early and late
quiescent phases. However, they strongly deviate from each other during
the ``onset'' phase in the middle. \label{fig:rescale-B} }

\end{figure}

The second set of simulations B1 -- B3 start with the initial condition
set by $E_{\perp}=1.4\times10^{-3}$. The initial profile and histogram
of $\sigma_{max}$ are shown in Figure \ref{fig:sigma_B}. The $\sigma_{max}$
profile and histogram are qualitatively similar to that of Case A.
The maximum $\sigma_{max}$ in the whole domain is approximately 7.4,
slightly higher than that of Case A. As we will see, although Case
A and Case B do not differ significantly in the initial $\sigma_{max}$
profiles, they lead to very different dynamical behavior at later
times.

Figure \ref{fig:CaseB-snapshot} shows a snapshot at $t=315$ from
Run B3, with an accompanying movie available online. From the time
histories of $J_{max}$ and $u_{max}$ shown in the upper left panel,
we can see that the system relaxes quiescently in the beginning (similar
to Case A) until $t\simeq260$. After that, both $J_{max}$ and $u_{max}$
increase rapidly by an order of magnitude and reach their peak magnitudes
at $t\simeq315$. As can be seen from the kinetic energy density,
footpoint drift and parallel voltage profiles in the snapshot at $t=315$,
the ``onset'' occurs in one sigmoid region where $\sigma_{max}$
is large. In this region, the footpoint drift speed reaches a peak
value $\simeq0.06$, the kinetic energy density increases by two orders
of magnitude compared to the quiescent phase. The maximum exponent
$\sigma_{max}$ increases to a peak value $\simeq10$ and the parallel
voltage increases by a factor of $\simeq5$ compared to the background
voltage before the onset, due to the formation of current filaments
in the onset region. After the system has released its magnetic free
energy, the evolution becomes quiescent again after $t\simeq450$.
We also note from the accompanying movie that during this ``onset''
phase the pattern of footpoint mapping undergoes a rapid qualitative
change, rather than a gradual deformation as during the quiescent
phases.

During the onset phase, a ribbon-like thin current sheet forms, as
shown by the $J=2$ isosurface in Figure \ref{fig:Isosurface}. We
also find signatures of magnetic reconnection near the current sheet.
Figure \ref{fig:reconnection} shows a 2D slice of current density
(color gradient), plasma flow (white arrows), and contours of flux
function $A$ (yellow solid lines) at the vicinity of the peak current
density. The flux function contours form a X-type geometry near the
current sheet, and the plasma flow exhibits the typical inflow and
outflow pattern of magnetic reconnection. The reconnection is asymmetric,
where the inflow from below is significantly stronger than that from
above. When the first current sheet starts to decay, a second current
sheet develops in the neighborhood (not shown).

Resistive relaxation of this initial condition has been carried out
for three different $\eta$. We again plot the rescaled kinetic energy
$E_{k}/\eta^{2}$ against the rescaled time $\eta t$, shown in Figure
\ref{fig:rescale-B}, as we did in case A. The three curves approximately
coincide with each other during the early and late quiescent phases.
However, they strongly deviate from each other during the onset phase
in the middle. Apparently, the early and late quiescent phases are
governed by resistive diffusion, with time scales $\sim1/\eta$. On
the other hand, the onset phase becomes narrower for smaller $\eta$
in the rescaled time $\eta t$, indicating that the time scale of
the onset phase is faster than $\sim1/\eta$. 

The peaks of total kinetic energy for Run B1 -- B3 are $2.9\times10^{-6}$,
$3.2\times10^{-6}$, and $2.7\times10^{-6}$, respectively. As such,
the kinetic energy at the onset phase does not strongly depend on
$\eta$. On the other hand, the current sheet that forms after the
onset becomes more intense the smaller the $\eta$ is. The peaks of
current density for Run B1 -- B3 are $7.3$, $18.4$, and $37.8$,
which approximately follow a $J_{peak}\sim1/\eta$ scaling. Because
we are not able to vary $\eta$ over a wide range due to the limitation
of resolution, these scaling relations should be considered as tentative
rather than definitive.

\section{Discussion\label{sec:Discussion}}

\subsection{Resistive Quasi-Static Evolution }

As we have observed from the simulation data, kinetic energy tends
to be high in regions where neighboring field lines strongly exponentiate
apart during the quiescent phase governed by resistive diffusion.
To gain a deeper understanding of this observation, here we consider
the analytic governing equations for resistive quasi-static evolution.
Suppose we start from a stable force-free equilibrium with a small
but nonzero $\eta$, the magnetic field will slowly diffuse, disturbing
exact force balance. In response, plasma flow will be induced. If
$\eta$ is sufficiently small, the plasma inertia and other terms
such as friction and viscosity are negligible, and the system will
evolve quasi-statically and remain close to a force-free equilibrium.
Therefore, the quasi-static evolution is governed by the condition
that the force-free condition is satisfied for all time, together
with the resistive induction equation, namely: 
\begin{equation}
\mathbf{B}\cdot\nabla J=0,\label{eq:qs1}
\end{equation}
\begin{equation}
\partial_{t}A=\mathbf{B}\cdot\nabla\phi-\eta J.\label{eq:qs2}
\end{equation}
By taking the time derivative of Equation (\ref{eq:qs1}) and using
Equation (\ref{eq:qs2}) to eliminate the time derivatives, we obtain
the following equation for $\phi$: 
\begin{equation}
\mathcal{L}\phi\equiv-\mathbf{B}\cdot\nabla\left(\nabla_{\perp}^{2}(\mathbf{B}\cdot\nabla\phi)\right)+\left[\mathbf{B}\cdot\nabla\phi,J\right]=-\eta\mathbf{B}\cdot\nabla\left(\nabla_{\perp}^{2}J\right).\label{eq:qs3}
\end{equation}
If the operator $\mathcal{L}$ is invertible subject to the boundary
condition $\phi|_{z=0}=\phi|_{z=L}=0$, we can in principle obtain
the stream function $\phi$ (and the flow $\mathbf{u}=\nabla_{\perp}\phi\times\mathbf{\hat{z}}$)
at each instant that will carry the system to another force-free equilibrium
at the next instant. 

It can be shown that the operator $\mathcal{L}$ is self-adjoint,
and is exactly the operator for the linear stability problem of the
system, which can be written as 
\begin{equation}
\mathcal{L}\phi=\omega^{2}\nabla_{\perp}^{2}\phi.\label{eq:linear_stability}
\end{equation}
Here a $\phi\sim e^{i\omega t}$ time dependence is assumed, and viscosity
and friction are neglected. Because the eigenfunctions of a self-adjoint
operator form a complete set, if the full set of eigenfunctions and
eigenvalues $\{\phi_{n},\omega_{n}^{2}\}$ is known, we can expand
the right hand side of Equation (\ref{eq:qs3}) with eigenfunctions
as $-\eta\mathbf{B}\cdot\nabla\left(\nabla_{\perp}^{2}J\right)=\sum_{n}a_{n}\nabla_{\perp}^{2}\phi_{n}$,
and the solution can formally be written as 
\begin{equation}
\phi=\sum_{n}\frac{a_{n}}{\omega_{n}^{2}}\phi_{n}.\label{eq:solution_qs}
\end{equation}

As we have noted earlier, $\nabla_{\perp}J$ tends to be large in
high-$\sigma$ regions even if $J$ itself may not be large, simply
because $J$ has to remain constant along a field line when the system
is in equilibrium. Roughly speaking, we may estimate $\nabla_{\perp}J\sim e^{\sigma}$
and $\nabla_{\perp}^{2}J\sim e^{2\sigma}$. Therefore, the right hand
side of Equation (\ref{eq:qs3}) in general will be large in high-$\sigma$
regions. When the operator $\mathcal{L}$ in Equation (\ref{eq:qs3})
is inverted to obtain the induced plasma flow, the flow is likely
to be large in high-$\sigma$ region as well. As a general rule, the
resistively induced flow is large in a high-$\sigma$ equilibrium,
which can be considered as enhanced resistive diffusion due to field
line exponentiation. As an example, the induced flow in Run A3 reach
a maximum $u_{max}\simeq1.8\times10^{-3}$, which is more than an
order of magnitude larger than a naive estimate from balancing characteristic
values of $\eta J$ and $uB$, which yields $u\simeq10^{-4}$. 

Another factor that affects the induced plasma flow is how close the
system is with respect to an ideal stability threshold. The formal
solution (\ref{eq:solution_qs}) diverges as the system approaches
an ideal stability threshold, i.e. when $\omega_{n}^{2}\to0^{+}$
for at least one eigenmode. If the system is close to an ideal stability
threshold, i.e. some $\omega_{n}^{2}$ are small, the induced flow
may become so large such that plasma inertia, friction, and viscosity
are no longer negligible, and the evolution will deviate from being
quasi-static. This may explain why the rescaled curves of four different
Runs in Case A as shown in Figure \ref{fig:rescale-A} deviate from
each other in the middle range of the plot, while they agree well
with each other in the early and late stages. As an overall trend
$\sigma_{max}$ tends to decrease as the system evolves due to resistive
decay of the magnetic field (see the movie of Case A), therefore the
induced flow energy should tend to decrease as well. However, the
resistive quasi-static evolution may actually bring the system closer
to a marginal stability threshold during the middle range. The fact
that both the maximum flow $u_{max}$ and the total kinetic energy
$E_{k}$ increase initially, even though $\sigma_{max}$ is decreasing,
lends support to this hypothesis. 

The formulation of quasi-static evolution here is similar to the one
derived by \citet{VanBallegooijen1985}{[}see also \citep{NgB1998}{]},
except that in his case the evolution is driven by footpoint motion,
instead of resistive diffusion. It is easy to generalize the formulation
to incorporate footpoint motion and resistive diffusion at the same
time. In the presence of footpoint motion prescribed by $\phi_{b}(\mathbf{x}_{\perp})$
and $\phi_{t}(\mathbf{x}_{\perp})$ at the bottom and top plates,
we have to solve Equation (\ref{eq:qs3}) subject to the boundary
conditions $\phi|_{z=0}=\phi_{b}$ and $\phi|_{z=L}=\phi_{t}$. Formally
this can be done by writing $\phi=\phi_{1}+\phi_{2}$, where $\phi_{2}$
is an arbitrary function that satisfies the boundary conditions. By
moving the contribution from $\phi_{2}$ to the right hand side of
Equation (\ref{eq:qs3}), we obtain the equation 
\begin{equation}
\mathcal{L}\phi_{1}=-\eta\mathbf{B}\cdot\nabla\left(\nabla_{\perp}^{2}J\right)-\mathcal{L}\phi_{2},\label{eq:qs_4}
\end{equation}
which can then be solved in the same way with the boundary condition
$\phi_{1}|_{z=0}=\phi_{1}|_{z=L}=0$ as before.

\subsection{Nature of the Onset in Case B}

\begin{figure}
\begin{centering}
\includegraphics[scale=0.8]{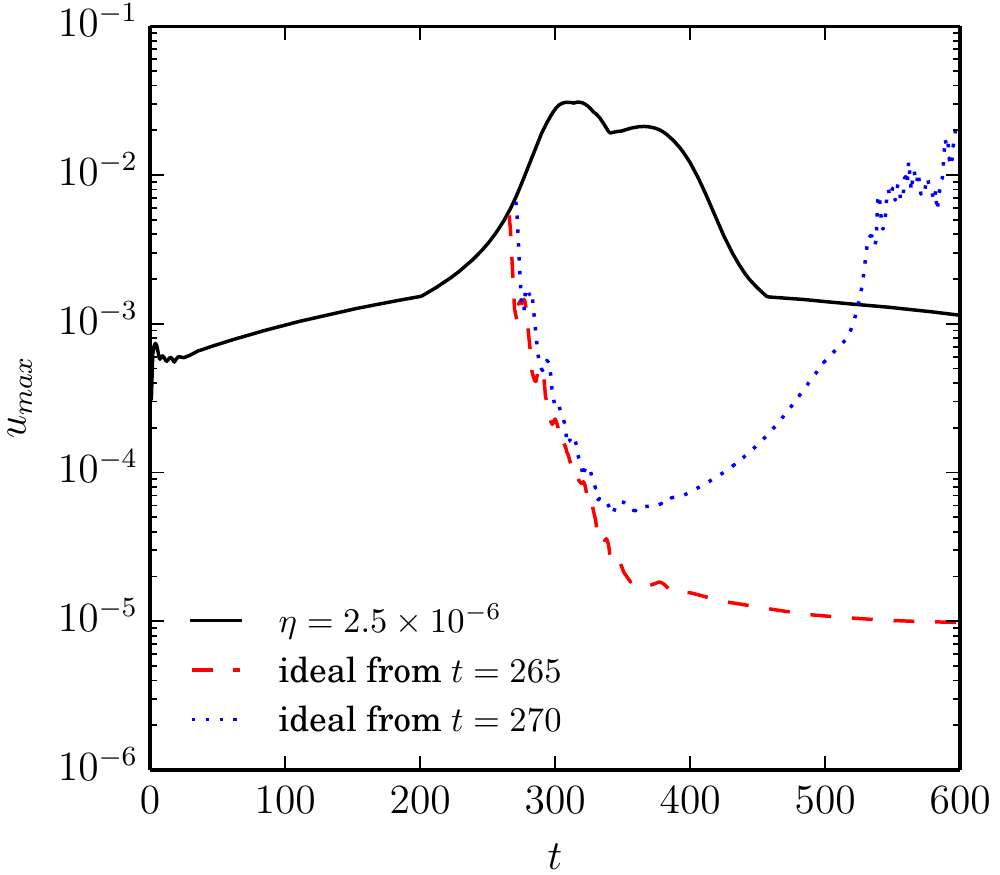}\includegraphics[scale=0.8]{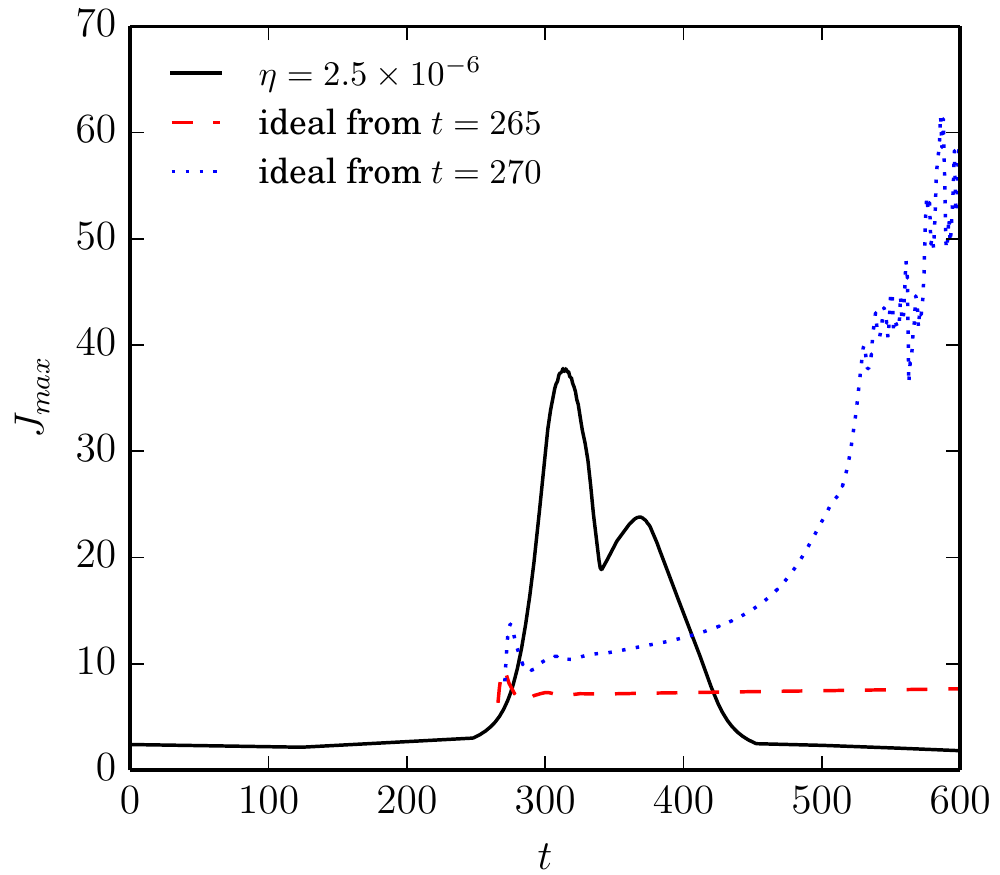}
\par\end{centering}

\protect\caption{Time histories of $u_{max}$ and $J_{max}$ from two restart runs
with $\eta$ set to zero. The black sold lines correspond to the original
Run B3; red dashed lines and blue dotted lines represent restart runs
from $t=265$ and $t=270$, respectively. \label{fig:restart} }

\end{figure}

We now address the important question of what causes the onset in
Case B. In previous discussion we suggest that resistive quasi-static
evolution may actually bring the system (as in Case A) closer to an
ideal stability threshold. Since Case B has a slightly higher initial
magnetic energy than Case A, it is possible that resistive quasi-static
evolution may eventually cause the system to pass through an ideal
stability threshold. To further examine this possibility, we take
the simulation B3 and restart at different times with $\eta$ set
to zero. Figure \ref{fig:restart} shows the time histories of $u_{max}$
and $J_{max}$ from two runs, restarted from $t=265$ and $t=270$,
respectively. For the restart run from $t=265$, the plasma flow is
quickly damped away by friction and viscosity, and the system relaxes
to a new equilibrium and stays there. On the contrary, for the restart
run from $t=270$, the plasma flow first decays and appears to be
settling down initially but eventually starts to increase at $t\simeq350$.
That leads to the formation of intense current sheets, sufficiently
intense that they eventually become under-resolved. These results
suggest that the system is still within an ideal stability threshold
at $t=265$, but has passed through the stability threshold at $t=270$.
{The current sheets formed after the onset of the
instability bear similarity with the ones caused by ideal coalescence
and kink instabilities \citep{LongcopeS1994,LongcopeS1994a,LionelloSEV1998,LionelloVEM1998,HoodBV2009}.
However, due to the lack of ignorable coordinates in the equilibrium,
analyzing the nature of the instability is a rather difficult task. }

Although these results suggest that the onset may be caused by crossing
an ideal MHD stability threshold, it is also possible that the system
becomes resistively unstable before it reaches the ideal stability
threshold. Formally we can analyze the resistive instability by assuming
that the background is not evolving, and linearizing Equations (\ref{eq:RMHD-momentum})
and (\ref{eq:RMHD-faraday}), which yields the following eigenvalue
problem: 
\begin{equation}
i\omega\Omega=\mathbf{B}\cdot\nabla\tilde{J}+[\tilde{A},J],\label{eq:RMHD-momentum-1}
\end{equation}
\begin{equation}
i\omega\tilde{A}=\mathbf{B}\cdot\nabla\phi+\eta\nabla_{\perp}^{2}\tilde{A}.\label{eq:RMHD-faraday-1}
\end{equation}
Here $\tilde{A}$ and $\tilde{J}$ denote the linear perturbations
of the flux function and current density, respectively; and we have
again neglected viscosity and friction. It has been previously found
that as the system just crosses the resistive stability threshold
in line-tied systems, the linear growth rate scales linearly with
$\eta$ \citep{DelzannoF2008,HuangZ2009}. The reason is that near
the stability threshold, the unstable mode is so slowly growing such
that the inertia term $i\omega\Omega$ in Equation (\ref{eq:RMHD-momentum-1})
is negligible. In this regime, the resistive instability is virtually
indistinguishable from resistive quasi-static diffusion. However,
as the system evolves further away from the stability threshold, the
growth rate is expected to scale as $\sim\eta^{\alpha}$, with $0<\alpha<1$,
and becomes distinguishable from resistive diffusion. To further examine
this possibility requires a clear separation between the resistive
diffusion time scale $\sim1/\eta$ and the resistive instability time
scale $\sim\eta^{-\alpha}$, which is beyond the parameter regimes
we are able to probe and is left to a future study.

\subsection{Measuring the ``Distance'' Between Field Line Mappings\label{sub:Measuring-the-Distance}}

\begin{figure}
\begin{centering}
\includegraphics[scale=0.9]{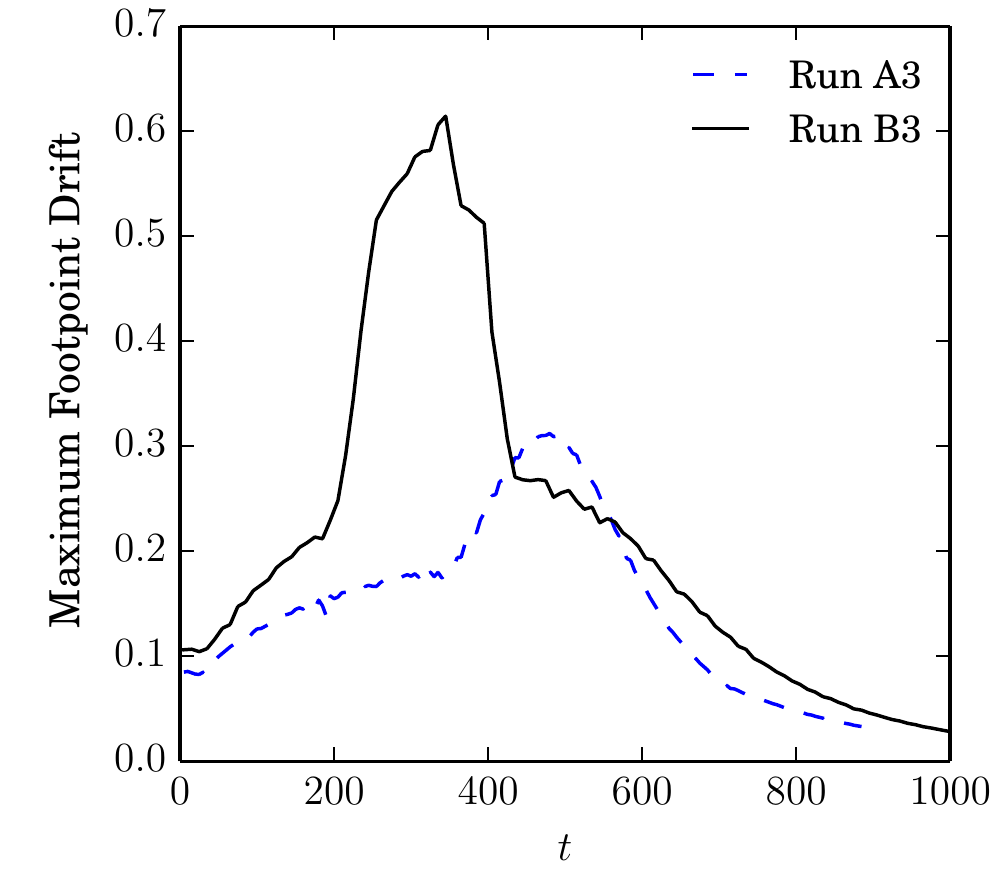}
\par\end{centering}

\protect\caption{Time histories of maximum footpoint drift between two consecutive
snapshots separated by one Alfv\'en transit time for Run A3 and Run
B3.\label{fig:Time-histories-of-max-drift}}
\end{figure}
\begin{figure}
\begin{centering}
\includegraphics[scale=0.9]{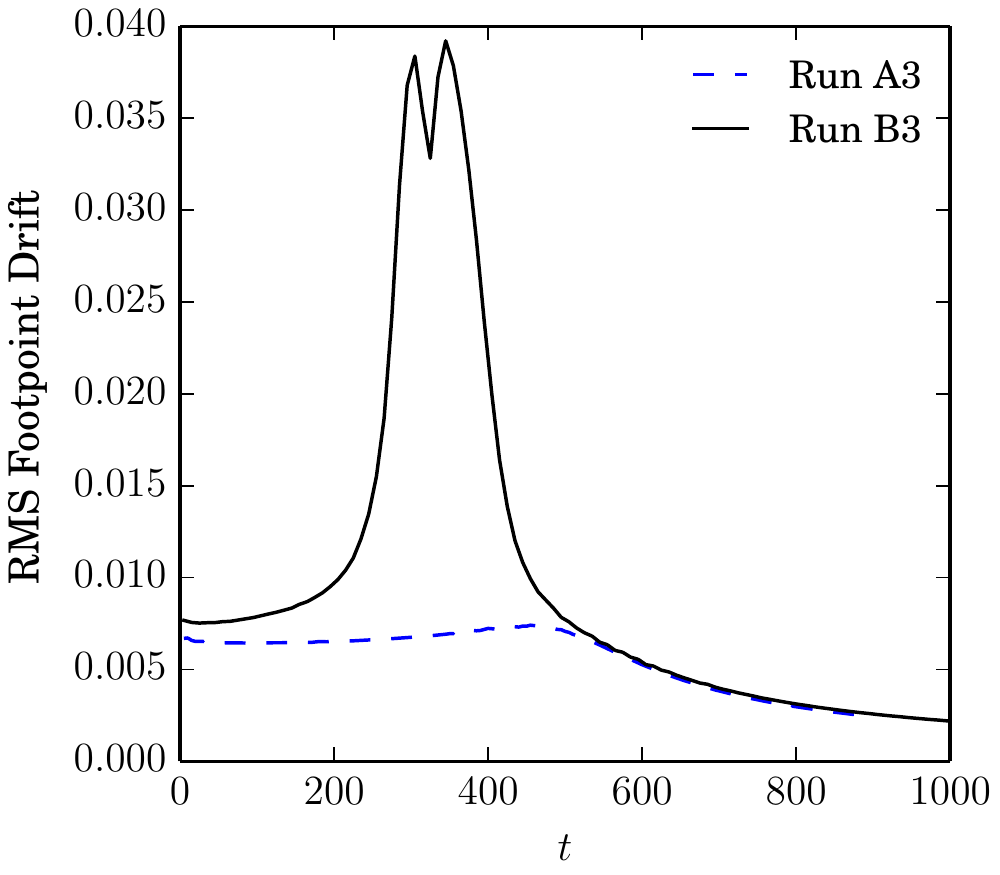}
\par\end{centering}

\protect\caption{Time histories of root mean square footpoint drift between two consecutive
snapshots separated by one Alfv\'en transit time for Run A3 and Run
B3.\label{fig:Time-histories-of-rms-drift}}
\end{figure}
\begin{figure}[h]
\begin{centering}
\includegraphics[scale=0.9]{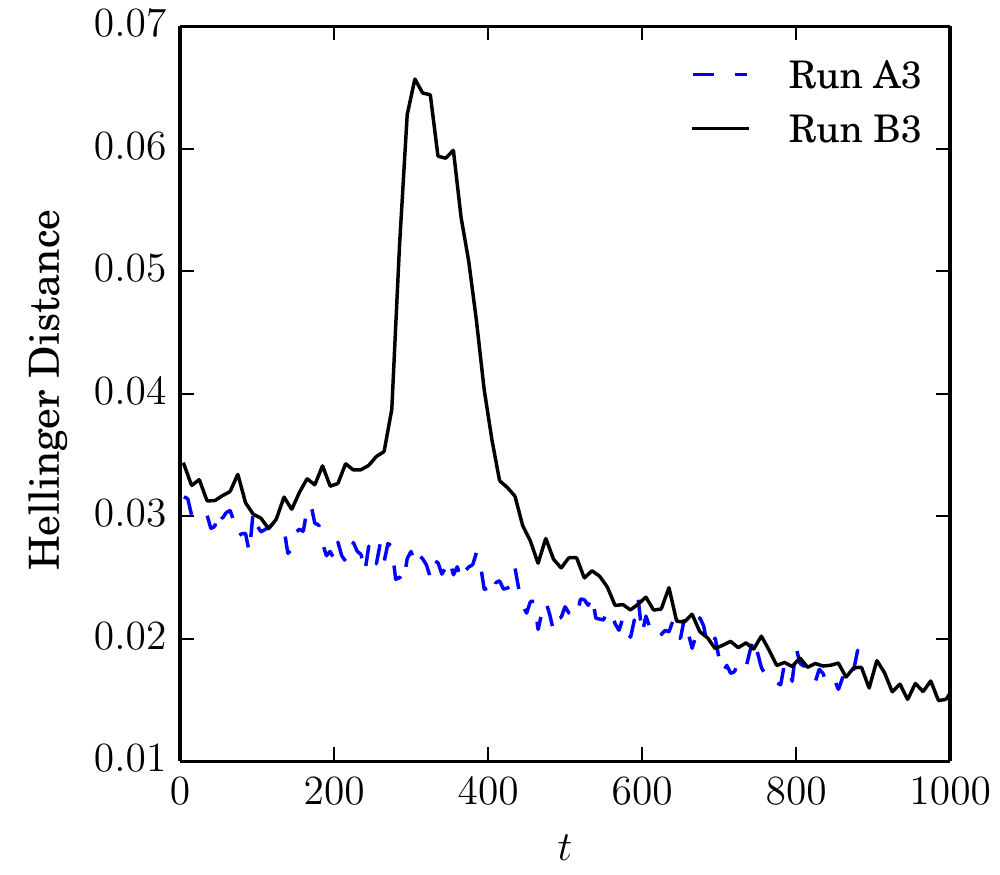}
\par\end{centering}

\protect\caption{Time histories of the Hellinger difference between two consecutive
snapshots of field line mapping separated by one Alfv\'en transit
time for Run A3 and Run B3. \label{fig:Hellinger} }
\end{figure}

In Section \ref{sec:Simulation-Results} we present two cases where
the evolutions of field line mapping follow qualitatively different
behavior. In the first case governed by resistive quasi-static evolution,
Run A3, the field line mapping appears to deform gradually; whereas
in the second case, Run B3, the field line mapping rapidly changes
to a qualitatively different pattern during the onset phase. Although
the difference between the two cases is rather obvious to the human
eye, how to translate this qualitative observation to a precise quantitative
definition of magnetic reconnection in three dimensions remains an
open question. To further address this issue, we ask the important
question: ``Given two consecutive snapshots of field line mappings,
how do we measure how different they are?''

A simple measure is the maximum footpoint drift between two consecutive
snapshots separated by one Alfv\'en transit time. However, as shown
by the time histories of maximum footpoint drift in Figure \ref{fig:Time-histories-of-max-drift},
the two cases A3 and B3 appear to be qualitatively similar. In both
cases, the maximum footpoint drift starts with a similar value $\simeq0.1$,
which gradually increases at later time, reaches a peak and then decays.
The peak value of maximum footpoint drift in case B3 is approximately
a factor of two higher than that in A3, which does not seem very significant.
The problem of taking the maximum value of footpoint drift is that,
it tends to overemphasize the difference, especially when the field
line mapping is highly distorted with large gradient, as in the present
cases. This can be appreciated by a simple one dimensional analogy.
Consider a one dimensional function with a strong gradient, e.g. $f_{1}(x)=\tanh(x/h)$
with $h\ll1$, and a slightly shifted function $f_{2}(x)=f_{1}(x-h)$.
Even though for all practical purposes $f_{1}(x)$ and $f_{2}(x)$
are very similar, the maximum value of $\left|f_{1}(x)-f_{2}(x)\right|$
is of order unity. In this case, instead of using the maximum value,
the $L^{2}$ norm of the difference 
\begin{equation}
\left|\left|f_{1}(x)-f_{2}(x)\right|\right|_{2}\equiv\left(\int\left|f_{1}(x)-f_{2}(x)\right|^{2}dx\right)^{1/2},\label{eq:L2}
\end{equation}
is a better measure of the ``distance'' between the two functions.

Likewise, given two field line mappings $\mathbf{x}_{1}(\mathbf{x}_{b})$
and $\mathbf{x}_{2}(\mathbf{x}_{b})$, where $\mathbf{x}_{b}$ represents
the coordinates at the bottom plate, we may measure the ``distance''
between them by the $L^{2}$ norm
\begin{equation}
\left|\left|\mathbf{x}_{1}(\mathbf{x}_{b})-\mathbf{x}_{2}(\mathbf{x}_{b})\right|\right|_{2}\equiv\left(\int\left|\mathbf{x}_{1}(\mathbf{x}_{b})-\mathbf{x}_{2}(\mathbf{x}_{b})\right|^{2}d^{2}x_{b}\right)^{1/2}.\label{eq:L2-1}
\end{equation}
In our diagnostics, we approximate the $L^{2}$ norm by the root mean
square (RMS) footpoint drift of the sampled field lines, and Figure
\ref{fig:Time-histories-of-rms-drift} shows the time histories of
the RMS footpoint drift, for the two cases. As we can see, the $L^{2}$
norm gives a sharp distinction between the two cases during the onset
phase of case B3.

Another possible way of measuring the distance between two mappings
is through coarse graining the field line connectivity. First we divide
the bottom plate and the top plate into $n\times n$ equal-sized cells
for each of them. The coarse-grained representation of a field line
mapping is the probability distribution $P=\{p_{ij}\}$, where $p_{ij}$
is the probability of a field line to connect the $i^{th}$ cell at
the bottom and the $j^{th}$ cell at the top. Here again we make use
of the doubly periodic boundary condition in defining field line connection
between the two plates, such that bottom plate of the simulation box
is exactly mapped to the top plate of the simulation box (see Figure
\ref{fig:Visualization-of-mapping}). The probability distribution
$\{p_{ij}\}$ satisfies the relation 
\begin{equation}
\sum_{ij}p_{ij}=1.\label{eq:normalization}
\end{equation}
After coarse-graining the field line mappings into probability distributions,
we can measure the distance between two mappings by a proper measure
of the distance between probability distributions. The motivation
behind the coarse-graining approach is that, if the field line mapping
deforms gradually in time, the probability distributions between two
consecutive snapshots should be very similar, even though footpoints
of individual field lines may drift a large distance.

Given two probability distributions $P=\{p_{ij}\}$ and $Q=\{q_{ij}\}$,
there are numerous standard ways to quantify the difference between
them {[}see, e.g. \citep{DasGupta2008}, Chapter 2, for an overview{]}.
Here we employ the Hellinger distance $H(P,Q)$, defined as 
\begin{equation}
H(P,Q)\equiv\frac{1}{\sqrt{2}}\sqrt{\sum_{ij}\left(\sqrt{p_{ij}}-\sqrt{q_{ij}}\right)^{2}}=\sqrt{1-\sum_{ij}\sqrt{p_{ij}q_{ij}}}.\label{eq:Hellinger}
\end{equation}
The Hellinger distance satisfies the inequality $0\le H(P,Q)\le1$.
We divide each of the two end plates into $10\times10$ cells, and
calculate the Hellinger distance between two consecutive field line
mappings separated by one Alfv\'en time by using the sampled field
lines to estimate the probability distributions. Figure \ref{fig:Hellinger}
shows the resulting time histories of Hellinger distance for the two
cases A3 and B3. As can be seen from the two curves in Figure \ref{fig:Hellinger},
the Hellinger distance is also very effective in distinguishing the
onset phase and the quasi-static phase.

\section{Summary and Conclusion\label{sec:Summary-and-Conclusion}}

In this study, we find two distinct phases in our simulations on resistive
relaxation of 3D force-free equilibria. In the phase governed by resistive
quasi-static evolution, it is found that kinetic energy tends to be
high, and field line mapping can change rapidly in regions where neighboring
field lines strongly exponentiate from each other (i.e. high-$\sigma$
regions). During this phase, the evolution time scale $\sim1/\eta$,
as in resistive diffusion. Whether we would call the change of field
line mapping ``reconnection'' is to some extent a matter of definition.
As we have seen, even though the mapping of individual field lines
can change rapidly, the overall pattern of the mapping appears to
deform gradually rather than to undergo rapid qualitative change.
Therefore, it may be more appropriate to call the rapid change of
field line connectivity and high induced plasma flow in high-$\sigma$
regions during this phase ``enhanced resistive diffusion'' rather
than ``reconnection''. On the other hand, we also find that in some
cases resistive quasi-static evolution can cause an ideally stable
initial force-free equilibrium to pass through a stability threshold,
leading to an ``onset'' phase. During this onset phase, intense
current filaments develop, and the field line mapping change rapidly
to a qualitatively different pattern on a timescale faster than $\sim1/\eta$.
The plasma flow exhibits signatures of inflow and outflow in the vicinities
of current filaments. When the same initial condition is relaxed with
different $\eta$, it is found that the current filaments become more
intense the smaller the $\eta$ is. The presence of all these hallmarks
of reconnection suggests that the change of field line connectivity
in this onset phase may be more properly designated as ``reconnection''.
Because of the different scaling relations with respect to $\eta$,
the difference between the two phases will become more distinct the
smaller the $\eta$ is. 

Our results illustrate a fundamental difference between 2D and 3D
reconnection. In 2D the breaking of magnetic field line connectivity
occurs in thin current sheets, therefore is much more closely coupled
with magnetic energy conversion and dissipation; whereas in 3D the
breaking of magnetic field line connectivity does not require a large
current density and, therefore, is not directly associated with a
large energy release. For this reason, a proper definition of reconnection
becomes a tricky issue. If we consider magnetic energy conversion
and dissipation as one of the hallmarks of magnetic reconnection,
then we must conclude that rapid change of field line connectivity
appears to be necessary, but insufficient as a signature of fast reconnection.
Nonetheless, our results show that the meaning of the ideal constraints
on magnetic field evolution is very subtle in 3D. Because of the exponential
sensitivity, strict conservation of field line mapping as dictated
by ideal MHD is essentially impossible in a generic, high-$\sigma$
3D magnetic field. On the other hand, the fact the field line mapping
can undergo gradual deformation even in the presence of a large apparent
field line drift also calls for quantification of approximate conservation
of ideal constraints other than direct measurement of individual footpoint
drift. The measures of $L_{2}$ norm and Hellinger distance in Section
\ref{sub:Measuring-the-Distance} represent our preliminary attempt
in this direction. 

The process of current sheet formation during the onset phase requires
further investigation. When the evolution deviates from the force-free
condition $\mathbf{B}\cdot\nabla J=0$ due to some instability, the
large gradient in $J$ implies the likelihood of developing a large
$dJ/d\ell$, where $d\ell$ is the differential distance along a field
line. A large $dJ/d\ell$ implies a large current across the magnetic
field, for otherwise the current density would not be divergence free.
The force associated with current across the field lines must be balanced
by inertia. This is just what is implied by Equation (\ref{eq:RMHD-momentum}):
a large $dJ/d\ell$ implies a large time derivative in the vorticity
of the flow $\Omega$. Although the reported simulations have sufficient
dissipation to prevent Alfv\'en waves from bouncing back and forth,
the relaxation of $dJ/d\ell$ back to zero occurs by Alfv\'en waves.
As shown in \citep{Boozer2014}, Alfv\'en wave propagation in a high-$\sigma$
region turns the part of $J$ that is not constant along a field line
into a current ribbon of exponentially high current density as well
as producing a vortex ribbon {[}see also \citep{SimilonS1989}{]}.
The current and vortex ribbon can be rapidly dissipated by resistivity
or viscosity. Moreover, this mechanism can take place without the
mediation of instabilities, as the large gradient in $J$ implies
that any small perturbation in $\mathbf{B}$ can potentially leads
to a large $dJ/dl$. Therefore, a general conclusion is that a high-$\sigma$
force-free equilibrium is very fragile and can hardly remain quiescent. 

From a broader perspective, the interesting finding that resistive
quasi-static evolution can lead to an onset of instability and subsequent
formation of thin current sheets may have profound implications on
Parker's ``topological dissipation'' scenario of coronal heating
\citep{Parker1972}. In this scenario, it is suggested that when the
magnetic field lines in solar coronal loops become entangled due to
footpoint motions on the photosphere, the ideal magnetostatic equilibrium
of the entangled field develops current singularities (i.e. tangential
discontinuities in magnetic field). In the presence of small but finite
resistivity, the current singularities can dissipate resistively and
heat the corona. Parker's proposal has since become the famous Parker
problem and has stimulated a substantial amount of debate that continues
to this day \citep{VanBallegooijen1985,ZweibelL1987,NgB1998,CraigS2005,Low2006a,JanseL2009,HuangBZ2009,HuangBZ2010,JanseL2010,AlyA2010,JanseLP2010,Low2010a,Craig2010,PontinH2012,Low2013}.
The Parker problem is usually posed as an ideal MHD problem, where
the key question is whether the minimum energy equilibrium for a given
magnetic topology, assuming ideal MHD constraints are exactly conserved,
contains tangential discontinuity or not. This formulation of the
Parker problem is mathematically precise, and to some degree carries
the connotation that formation of current sheets is purely an ideal
MHD process, whereas the role of resistivity is to smooth out the
singularity, to facilitate reconnection, and to dissipate the free
energy. However, as neighboring field line exponentiation becomes
significant as a natural consequence of continuous footpoint driving,
conservation of ideal MHD constraints becomes subtle and may become
overly restrictive. Furthermore, our results show that a small resistivity
can cause an otherwise ideally stable equilibrium to break loose and
subsequently develop much more intense current filaments. Therefore,
it appears that resistivity can play an active, albeit subtle role
in the formation of current sheets. Consequently, the physically relevant
question may not be the ideal problem, but the resistive problem in
the limit of very small $\eta$. This viewpoint has previously been
emphasized by \citet{BhattacharjeeW1991} and now corroborated by
the present study. A similar onset of instability has been reported
in recent studies on resistive relaxation of braided force-free coronal
loops \citep{Wilmot-SmithPH2010,PontinWHG2011}. However, the onset
time is approximately independent of $\eta$ in the reported simulations,
whereas the onset time $\sim1/\eta$ in the present study. Therefore,
the instabilities in the two types of studies are likely to be qualitatively
different.

{The enhanced resistive diffusion reported in this
work is conceptually different from the mechanism of breaking ideal
MHD frozen-in condition via turbulent Richardson advection as proposed
by \citet{EyinkVLAKBBMS2013}. In their scenario, the spontaneously
stochastic trajectories caused by turbulent flow bring together field
lines from distances far apart, which, as Eyink et al. argued, results
in breakdown of the frozen-in condition even in the limit $\eta\to0$.
On the other hand, the enhanced resistive diffusion in the present
work is caused by the stochasticity in magnetic field line mapping.
The plasma flow does not become turbulent in either the quiescent
or onset phases. Therefore, Alfv\'en's frozen-in theorem is expected
to be satisfied as $\eta\to0$.}

Finally, we remark on some limitations of the present study. The initial
conditions employed in this study are constructed by integrating $\mathbf{B}\cdot\nabla J=0$
from a given flux function $A_{0}$ at the mid-plane of the simulation
box; these initial conditions may not be representative of the generic
force-free equilibria that arise more naturally, e.g. by imposing
footpoint motion at the boundaries as in Parker's model. The exponent
$\sigma$ in this work is rather modest due to the limitation of numerical
resolution, while high-$\sigma$ regions concentrate in layers. In
astrophysical or space plasmas, the exponent $\sigma$ is expected
to be significantly higher, and high-$\sigma$ regions could be more
volume-filling. Other physical effects known to be important to reconnection,
e.g. Hall effect, electron pressure and inertia, are also not included.
The conclusions of this study should be further examined under those
conditions.

\acknowledgements{YMH would like to dedicate his work on this paper to Prof. Dalton
Schnack. Advice on 3D visualization from Dr. Liwei Lin and Burlen
Loring are highly appreciated. {We also thank the
anonymous referee for constructive comments to improve the presentation.
This work is facilitated by the Max-Planck/Princeton for Plasma Physics
and supported by the Department of Energy, Grant No. DE-FG02-07ER46372,
under the auspice of the Center for Integrated Computation and Analysis
of Reconnection and Turbulence (CICART), the National Science Foundation,
Grant No. PHY-0215581 (PFC: Center for Magnetic Self-Organization
in Laboratory and Astrophysical Plasmas), NASA Grant Nos. NNX09AJ86G
and NNX10AC04G, and NSF Grant Nos. ATM-0802727, ATM-090315, AGS-1338944,
and AGS-0962698. Computations were performed on facilities at National
Energy Research Scientific Computing Center. }}

\appendix

\section{{Construction of Force-Free Equilibrium\label{sec:Construction-of-Force-Free}}}

{The initial force-free equilibrium for simulation
is constructed by integrating 
\begin{equation}
\mathbf{B}\cdot\nabla J=\partial_{z}J+\left[A,J\right]=0\label{eq:force-free}
\end{equation}
along the $z$ direction from a prescribed $A=A_{0}(\mathbf{x}_{\perp})$
at $z=0$. Although Equation (\ref{eq:force-free}) can in principle
be integrated with any standard numerical scheme, because the constructed
equilibrium is usually very delicate due to high field line exponentiation,
it is important to employ exactly the same discretization scheme for
both the simulation code and the integration; otherwise, the discrepancy
in discretization errors due to different numerical schemes may result
in non-negligible unbalanced force when the constructed equilibrium
is used as an initial condition for simulation. }

{The simulation code employs staggered grids for variables
$\phi$ and $A$ along the $z$ direction. That means that $\mathbf{B}\cdot\nabla J$
is evaluated at the middle point between two adjacent grid points
where $A$ resides. Specifically, the discretized representation for
$\mathbf{B}\cdot\nabla J$ between the $i$\textsuperscript{th} and
$(i+1)$\textsuperscript{th} grid points is }

{
\begin{equation}
\left(\mathbf{B}\cdot\nabla J\right)_{i+1/2}=\frac{J_{i+1}-J_{i}}{\Delta z}+\left[\frac{A_{i+1}+A_{i}}{2},\frac{J_{i+1}+J_{i}}{2}\right],\label{eq:discretization}
\end{equation}
where $\Delta z$ is the grid size along $z$; the Poisson bracket
is calculated by a standard pseudospectral method, i.e. derivatives
are carried out in the Fourier space and products are calculated in
the real space. When the discretized representation (\ref{eq:discretization})
is used in Equation (\ref{eq:force-free}), it becomes an implicit
scheme that $A_{i+1}$ can be determined by inverting a nonlinear
equation if $A_{i}$ is known. This nonlinear inversion is solved
iteratively until a convergence criterion is satisfied. In this work,
the convergence criterion is set to be $\int\left(\mathbf{B}\cdot\nabla J\right)_{i+1/2}^{2}d^{2}x<10^{-25}.$ }

{The corresponding problem in full MHD instead of
RMHD is constructing a force-free equilibrium from all three components
of $\mathbf{B}$ given on a plane. This is an important problem in
solar physics due to the great interest in extrapolating the coronal
magnetic field from the photospheric magnetograms {[}see, e.g. \citep{WiegelmannS2012}
for a recent review and references therein{]}. A straightforward generalization
of the procedure here to full MHD is the well-known upward integration
method \citep{Nakagawa1974,WuSCHG1990}, which integrates $\nabla\times\mathbf{B}=\alpha\mathbf{B}$
and $\nabla\cdot\mathbf{B}=0$ simultaneously from the vector }\textbf{{$\mathbf{B}$}}{{}
given at a plane (assumed to be the $z=0$ plane in the following
discussion). First the variable $\alpha$ is determined by $z$ component
of $\nabla\times\mathbf{B}=\alpha\mathbf{B}$: }

{
\begin{equation}
\partial_{x}B_{y}-\partial_{y}B_{x}=\alpha B_{z}.\label{eq:alpha}
\end{equation}
The remaining two components of $\nabla\times\mathbf{B}=\alpha\mathbf{B}$
then give the $z$ derivatives of $B_{x}$ and $B_{y}$: 
\begin{equation}
\partial_{z}B_{x}=\alpha B_{y}+\partial_{x}B_{z},\label{eq:dzBx}
\end{equation}
\begin{equation}
\partial_{z}B_{y}=-\alpha B_{x}-\partial_{y}B_{z}.\label{eq:dzBy}
\end{equation}
Finally, the $z$ derivative of $B_{z}$ is obtained from the condition
$\nabla\cdot\mathbf{B}=0$:}

{
\begin{equation}
\partial_{z}B_{z}=-\partial_{x}B_{x}-\partial_{y}B_{y}.\label{eq:dzBz}
\end{equation}
Equations (\ref{eq:alpha}) -- (\ref{eq:dzBz}) can be employed to
integrate all three components of $\mathbf{B}$ to the next constant-$z$
plane and the whole procedure is then repeated. }

{However, the upward integration method in full MHD
is known to be mathematically ill-posed and numerically unstable {[}see,
e.g. the discussions in \citep{LowL1990,AmariALBM1997,DemoulinHMP1997}{]}.
In particular, difficulties may arise near photospheric polarity inversion
lines where $B_{z}=0$ (which are present in most solar applications),
as Equation (\ref{eq:alpha}) requires that the left hand side $\partial_{x}B_{y}-\partial_{y}B_{x}$
must vanish as well. Therefore, the three components of $\mathbf{B}$
at $z=0$ cannot be arbitrarily prescribed but must satisfy the above
mentioned constraint. Although $\alpha$ can in principle be obtained
by applying l'H\^ospital's rule if the constraint is satisfied \citep{CupermanDS1991},
any arbitrary small errors in $\mathbf{B}$ can violate the constraint
and render the boundary condition incompatible with the force-free
condition. Another drawback of the upward integration method for coronal
magnetic field extrapolation is that the desirable asymptotic condition
that the magnetic field vanishes at infinity cannot be guaranteed
because boundary conditions are only imposed at $z=0$. Direct integration
of Equations (\ref{eq:alpha}) -- (\ref{eq:dzBz}) usually leads to
exponential growth of the magnetic field away from the boundary, and
the solution is very sensitive to tiny changes of the boundary conditions.
Various techniques have been proposed to regularize the method \citep{CupermanOS1990a,DemoulinCS1992},
but cannot be considered as fully successful. }

{In the context of the present study, however, the
above mentioned limitations of the upward integration method are largely
alleviated. The assumption of a strong $B_{z}$ ensures that Equation
(\ref{eq:alpha}) is well-conditioned. Furthermore, the force-free
equations are only integrated over a finite distance, therefore the
asymptotic behavior of $\mathbf{B}$ is not a concern. It remains
an interesting open question whether a force-free equilibrium can
be constructed for full MHD by a method similar to the one presented
here under these conditions.}


\begin{thebibliography}{}
\expandafter\ifx\csname natexlab\endcsname\relax\def\natexlab#1{#1}\fi

\bibitem[{Aly \& Amari(2010)}]{AlyA2010}
Aly, J.~J., \& Amari, T. 2010, Astrophys. J. Lett., 709, L99

\bibitem[{Amari {et~al.}(1997)Amari, Aly, Luciani, Boulmezaoud, \&
  Mikic}]{AmariALBM1997}
Amari, T., Aly, J.~J., Luciani, J.~F., Boulmezaoud, T.~Z., \& Mikic, Z. 1997,
  Solar Physics, 174, 129

\bibitem[{Bhattacharjee \& Wang(1991)}]{BhattacharjeeW1991}
Bhattacharjee, A., \& Wang, X. 1991, Astrophys. J., 372, 321

\bibitem[{Biskamp(2000)}]{Biskamp2000}
Biskamp, D. 2000, Magnetic Reconnection in Plasmas (Cambridge University Press)

\bibitem[{Boozer(2012{\natexlab{a}})}]{Boozer2012a}
Boozer, A.~H. 2012{\natexlab{a}}, Phys. Plasmas, 19, 092902

\bibitem[{Boozer(2012{\natexlab{b}})}]{Boozer2012}
---. 2012{\natexlab{b}}, Phys. Plasmas, 19, 112901

\bibitem[{Boozer(2013)}]{Boozer2013}
---. 2013, Phys. Plasmas, 20, 032903

\bibitem[{Boozer(2014)}]{Boozer2014}
---. 2014, Phys. Plasmas, in press

\bibitem[{Craig(2010)}]{Craig2010}
Craig, I. J.~D. 2010, Solar Physics, 266, 293

\bibitem[{Craig \& Sneyd(2005)}]{CraigS2005}
Craig, I. J.~D., \& Sneyd, A.~D. 2005, Solar Physics, 232, 41

\bibitem[{Cuperman {et~al.}(1991)Cuperman, D\'emoulin, \&
  Semel}]{CupermanDS1991}
Cuperman, S., D\'emoulin, P., \& Semel, M. 1991, Astron. Astrophys., 245, 285

\bibitem[{Cuperman {et~al.}(1990)Cuperman, Ofman, \& Semel}]{CupermanOS1990a}
Cuperman, S., Ofman, L., \& Semel, M. 1990, Astron. Astrophys., 230, 193

\bibitem[{DasGupta(2008)}]{DasGupta2008}
DasGupta, A. 2008, Asymptotic Theory of Statistics and Probability (New York:
  Springer Science+Business Media, LLC)

\bibitem[{Delzanno \& Finn(2008)}]{DelzannoF2008}
Delzanno, G.~L., \& Finn, J.~M. 2008, Phys. Plasmas, 15, 032904

\bibitem[{D\'emoulin {et~al.}(1992)D\'emoulin, Cuperman, \&
  Semel}]{DemoulinCS1992}
D\'emoulin, P., Cuperman, S., \& Semel, M. 1992, Astron. Astrophys., 263, 351

\bibitem[{D\'emoulin {et~al.}(1997)D\'emoulin, H\'enoux, Mandrini, \&
  Priest}]{DemoulinHMP1997}
D\'emoulin, P., H\'enoux, J.~C., Mandrini, C.~H., \& Priest, E.~R. 1997, Solar
  Physics, 174, 73

\bibitem[{D\'emoulin {et~al.}(1996)D\'emoulin, H\'enoux, Priest, \&
  Mandrini}]{DemoulinHPM1996}
D\'emoulin, P., H\'enoux, J.~C., Priest, E.~R., \& Mandrini, C.~H. 1996,
  Astron. Astrophys., 308, 643

\bibitem[{Dmitruk \& G\'omez(1999)}]{DmitrukG1999}
Dmitruk, P., \& G\'omez, D.~O. 1999, Astrophys. J., 527, L63

\bibitem[{Dmitruk {et~al.}(2003)Dmitruk, G\'omez, \& Matthaeus}]{DmitrukGM2003}
Dmitruk, P., G\'omez, D.~O., \& Matthaeus, W.~H. 2003, Phys. Plasmas, 10, 3584

\bibitem[{Eyink {et~al.}(2013)Eyink, Vishniac, Lalescu, Aluie, Kanov, B\"urger,
  Burns, Meneveau, \& Szalay}]{EyinkVLAKBBMS2013}
Eyink, G., Vishniac, E., Lalescu, C., {et~al.} 2013, Nature, 497, 466

\bibitem[{Finn {et~al.}(2014)Finn, Billey, Daughton, \& Zweibel}]{FinnBDZ2014}
Finn, J.~M., Billey, Z., Daughton, W., \& Zweibel, E. 2014, Plasma Phys.
  Control. Fusion, 56, 064013

\bibitem[{Gekelman {et~al.}(2014)Gekelman, Van~Compernolle, DeHass, \&
  Vincena}]{GekelmanVDV2014}
Gekelman, W., Van~Compernolle, B., DeHass, T., \& Vincena, S. 2014, Plasma
  Phys. Control. Fusion, 56, 064002

\bibitem[{Greene(1993)}]{Greene1993}
Greene, J.~M. 1993, Phys. Fluids B, 5, 2355

\bibitem[{Hesse {et~al.}(2005)Hesse, Forbes, \& Birn}]{HesseFB2005}
Hesse, M., Forbes, T.~G., \& Birn, J. 2005, Astrophys. J., 631, 1227

\bibitem[{Hesse \& Schindler(1988)}]{HesseS1988}
Hesse, M., \& Schindler, K. 1988, J. Geophy. Res., 93, 5559

\bibitem[{Hood {et~al.}(2009)Hood, Browning, \& Van~der Linden}]{HoodBV2009}
Hood, A.~W., Browning, P.~K., \& Van~der Linden, R. A.~M. 2009, Astron.
  Astrophys., 506, 913

\bibitem[{Huang {et~al.}(2009)Huang, Bhattacharjee, \& Zweibel}]{HuangBZ2009}
Huang, Y.-M., Bhattacharjee, A., \& Zweibel, E.~G. 2009, Astrophys. J. Lett.,
  699, L144

\bibitem[{Huang {et~al.}(2010)Huang, Bhattacharjee, \& Zweibel}]{HuangBZ2010}
---. 2010, Phys. Plasmas, 17, 055707

\bibitem[{Huang \& Zweibel(2009)}]{HuangZ2009}
Huang, Y.-M., \& Zweibel, E.~G. 2009, Phys. Plasmas, 16, 042102

\bibitem[{Janse \& Low(2009)}]{JanseL2009}
Janse, A.~M., \& Low, B.~C. 2009, Astrophys. J., 690, 1089

\bibitem[{Janse \& Low(2010)}]{JanseL2010}
---. 2010, Astrophys. J., 722, 1844

\bibitem[{Janse {et~al.}(2010)Janse, Low, \& Parker}]{JanseLP2010}
Janse, A.~M., Low, B.~C., \& Parker, E.~N. 2010, Phys. Plasmas, 17, 092901

\bibitem[{Kadomtsev \& Pogutse(1974)}]{KadomtsevP1974}
Kadomtsev, B.~B., \& Pogutse, O.~P. 1974, Sov. Phys. JETP, 38, 283

\bibitem[{Lawrence \& Gekelman(2009)}]{LawrenceG2009}
Lawrence, E.~E., \& Gekelman, W. 2009, PRL, 103, 105002

\bibitem[{Lionello {et~al.}(1998{\natexlab{a}})Lionello, Schnack, Einaudi, \&
  Velli}]{LionelloSEV1998}
Lionello, R., Schnack, D.~D., Einaudi, G., \& Velli, M. 1998{\natexlab{a}},
  Phys. Plasmas, 54, 3722

\bibitem[{Lionello {et~al.}(1998{\natexlab{b}})Lionello, Velli, Einaudi, \&
  Miki\'c}]{LionelloVEM1998}
Lionello, R., Velli, M., Einaudi, G., \& Miki\'c, Z. 1998{\natexlab{b}},
  Astrophys. J., 494, 840

\bibitem[{Longcope \& Strauss(1994{\natexlab{a}})}]{LongcopeS1994a}
Longcope, D.~W., \& Strauss, H.~R. 1994{\natexlab{a}}, Astrophys. J., 437, 851

\bibitem[{Longcope \& Strauss(1994{\natexlab{b}})}]{LongcopeS1994}
---. 1994{\natexlab{b}}, Astrophys. J., 426, 742

\bibitem[{Longcope \& Sudan(1994)}]{LongcopeS1994b}
Longcope, D.~W., \& Sudan, R.~N. 1994, Astrophys. J., 437, 491

\bibitem[{Low(2006)}]{Low2006a}
Low, B.~C. 2006, Astrophys. J., 649, 1064

\bibitem[{Low(2010)}]{Low2010a}
---. 2010, Solar Physics, 266, 277

\bibitem[{Low(2013)}]{Low2013}
---. 2013, Astrophys. J., 768, 7

\bibitem[{Low \& Lou(1990)}]{LowL1990}
Low, B.~C., \& Lou, Y.~Q. 1990, Astrophys. J., 352, 343

\bibitem[{Nakagawa(1974)}]{Nakagawa1974}
Nakagawa, Y. 1974, Astrophys. J., 190, 437

\bibitem[{Ng \& Bhattacharjee(1998)}]{NgB1998}
Ng, C.~S., \& Bhattacharjee, A. 1998, Phys. Plasmas, 5, 4028

\bibitem[{Ng \& Bhattacharjee(2008)}]{NgB2008}
---. 2008, Astrophys. J., 675, 899

\bibitem[{Ng {et~al.}(2012)Ng, Lin, \& Bhattacharjee}]{NgLB2012}
Ng, C.~S., Lin, L., \& Bhattacharjee, A. 2012, Astrophys. J., 747, 109

\bibitem[{Parker(1972)}]{Parker1972}
Parker, E.~N. 1972, Astrophys. J., 174, 499

\bibitem[{Pontin(2011)}]{Pontin2011}
Pontin, D.~I. 2011, Adv. Space Res., 47, 1508

\bibitem[{Pontin \& Huang(2012)}]{PontinH2012}
Pontin, D.~I., \& Huang, Y.-M. 2012, Astrophys. J., 756, 7

\bibitem[{Pontin {et~al.}(2011)Pontin, Wilmot-Smith, Hornig, \&
  Galsgaard}]{PontinWHG2011}
Pontin, D.~I., Wilmot-Smith, A.~L., Hornig, G., \& Galsgaard, K. 2011, Astron.
  Astrophys., 525, A57

\bibitem[{Priest \& D\'emoulin(1995)}]{PriestD1995}
Priest, E.~R., \& D\'emoulin, P. 1995, J. Geophy. Res., 100, 23443

\bibitem[{Priest \& Forbes(2000)}]{PriestF2000}
Priest, E.~R., \& Forbes, T. 2000, Magnetic reconnection : {MHD} theory and
  applications (Cambridge University Press)

\bibitem[{Priest {et~al.}(2003)Priest, Hornig, \& Pontin}]{PriestHP2003}
Priest, E.~R., Hornig, G., \& Pontin, D.~I. 2003, Journal of Geophysical
  Research, 108, 1285

\bibitem[{Rappazzo \& Parker(2013)}]{RappazzoP2013}
Rappazzo, A.~F., \& Parker, E.~N. 2013, Astrophys. J. Lett., 773, L2

\bibitem[{Rappazzo {et~al.}(2007)Rappazzo, Velli, Einaudi, \&
  Dahlburg}]{RappazzoVED2007}
Rappazzo, A.~F., Velli, M., Einaudi, G., \& Dahlburg, R.~B. 2007, Astrophys.
  J., 657, L47

\bibitem[{Rappazzo {et~al.}(2008)Rappazzo, Velli, Einaudi, \&
  Dahlburg}]{RappazzoVED2008}
---. 2008, Astrophys. J., 677, 1348

\bibitem[{Richardson \& Finn(2012)}]{RichardsonF2012}
Richardson, A.~S., \& Finn, J.~M. 2012, Commun. Nonlinear Sci. Numer. Simulat.,
  17, 2132

\bibitem[{Schindler {et~al.}(1988)Schindler, Hesse, \& Birn}]{SchindlerHB1988}
Schindler, K., Hesse, M., \& Birn, J. 1988, J. Geophy. Res., 93, 5547

\bibitem[{Schnack {et~al.}(1986)Schnack, Barnes, Miki\'c, Harned, Caramana, \&
  Nebel}]{SchnackBMHCN1986}
Schnack, D.~D., Barnes, D.~C., Miki\'c, Z., {et~al.} 1986, Computer Physics
  Communications, 43, 17

\bibitem[{Similon \& Sudan(1989)}]{SimilonS1989}
Similon, P.~L., \& Sudan, R.~N. 1989, Astrophys. J., 336, 442

\bibitem[{Strauss(1976)}]{Strauss1976}
Strauss, H.~R. 1976, Phys. Fluids, 19, 134

\bibitem[{Strauss \& Otani(1988)}]{StraussO1988}
Strauss, H.~R., \& Otani, N.~F. 1988, Astrophys. J., 326, 418

\bibitem[{Titov(2007)}]{Titov2007}
Titov, V.~S. 2007, Astrophys. J., 660, 863

\bibitem[{Trefethen \& Bau(1997)}]{TrefethenB1997}
Trefethen, L.~N., \& Bau, III, D. 1997, Numerical Linear Algebra (SIAM
  Philadelphia)

\bibitem[{van Ballegooijen(1985)}]{VanBallegooijen1985}
van Ballegooijen, A.~A. 1985, Astrophys. J., 298, 421

\bibitem[{Wiegelmann \& Sakurai(2012)}]{WiegelmannS2012}
Wiegelmann, T., \& Sakurai, T. 2012, Living Reviews in Solar Physics, 9,
  doi:10.12942/lrsp-2012-5

\bibitem[{Wilmot-Smith {et~al.}(2010)Wilmot-Smith, Pontin, \&
  Hornig}]{Wilmot-SmithPH2010}
Wilmot-Smith, A.~L., Pontin, D.~I., \& Hornig, G. 2010, Astron. Astrophys.,
  516, A5

\bibitem[{Wu {et~al.}(1990)Wu, Sun, Chang, Hagyard, \& Gary}]{WuSCHG1990}
Wu, S.~T., Sun, M.~T., Chang, H.~M., Hagyard, M.~J., \& Gary, G.~A. 1990,
  Astrophys. J., 362, 698

\bibitem[{Yamada {et~al.}(2010)Yamada, Kulsrud, \& Ji}]{YamadaKJ2010}
Yamada, M., Kulsrud, R., \& Ji, H. 2010, Rev. Mod. Phys., 82, 603

\bibitem[{Zweibel \& Li(1987)}]{ZweibelL1987}
Zweibel, E.~G., \& Li, H.-S. 1987, Astrophys. J., 312, 423

\bibitem[{Zweibel \& Yamada(2009)}]{ZweibelY2009}
Zweibel, E.~G., \& Yamada, M. 2009, Annu. Rev. Astron. Astrophys., 47, 291

\end{thebibliography}
\end{document}